\documentclass[fleqn,usenatbib]{mnras}

\usepackage{newtxtext,newtxmath}

\usepackage[T1]{fontenc}

\DeclareRobustCommand{\VAN}[3]{#2}
\let\VANthebibliography\thebibliography
\def\thebibliography{\DeclareRobustCommand{\VAN}[3]{##3}\VANthebibliography}

\usepackage{graphicx}	
\usepackage{amsmath}	
\usepackage{footnote}
\usepackage{comment}
\usepackage{threeparttable}
\usepackage{ragged2e}
\usepackage[normalem]{ulem}

\defcitealias{williams22}{I}
\newcommand*{\paperone}{Paper~\citetalias{williams22}}

\title[ALMA observations of G19.01--0.03 -- {\sc \textit{II}}.]{ALMA observations of the Extended Green Object G19.01--0.03: \\{\sc II}. A massive protostar with typical chemical abundances surrounded by four low-mass prestellar core candidates} 

\author[G. M. Williams et al.]{G. M. Williams,$^{1}$\thanks{E-mail: G.M.Williams@leeds.ac.uk}
C. J. Cyganowski,$^{2}$
C. L. Brogan,$^{3}$
T. R. Hunter,$^{3}$
P. Nazari$^{4}$
and R. J. Smith$^{5,2}$
\\
$^{1}$School of Physics \& Astronomy, The University of Leeds, Leeds, LS2 9JT, UK\\$^{2}$Scottish Universities Physics Alliance (SUPA), School of Physics and Astronomy, University of St Andrews, North Haugh, St Andrews, KY16 9SS, UK\\$^{3}$National Radio Astronomy Observatory (NRAO), 520 Edgemont Rd, Charlottesville, VA 22903, USA\\$^{4}$Leiden Observatory, Leiden University, PO Box 9513, 2300 RA Leiden, NL\\$^{5}$Jodrell Bank Centre for Astrophysics, Department of Physics and Astronomy, University of Manchester, Oxford Road, Manchester, M13 9PL, UK
}

\date{Accepted XXX. Received YYY; in original form ZZZ}

\pubyear{2023}

\begin{document}
\label{firstpage}
\pagerange{\pageref{firstpage}--\pageref{lastpage}}
\maketitle

\begin{abstract}
We present a study of the physical and chemical properties of the Extended Green Object (EGO) G19.01$-$0.03 using sub-arcsecond angular resolution Atacama Large Millimeter/submillimeter Array (ALMA) $1.05$\,mm and Karl G. Jansky Very Large Array (VLA) $1.21$\,cm data. G19.01$-$0.03 MM1, the millimetre source associated with the central massive young stellar object (MYSO), appeared isolated and potentially chemically young in previous Submillimeter Array observations.
In our $\sim0.4\arcsec$-resolution ALMA data, MM1 has four low-mass millimetre companions within $0.12$\,pc, all lacking maser or outflow emission, indicating they may be prestellar cores. With a rich ALMA spectrum full of complex organic molecules, 
MM1 does not appear chemically young, but has molecular abundances typical of high-mass hot cores in the literature.
At the $1.05$\,mm continuum peak of MM1, $\mathrm{N}(\mathrm{CH}_{3}\mathrm{OH})=(2.22\pm0.01)\times10^{18}$\,cm$^{-2}$ and $T_{\mathrm{ex}} = 162.7\substack{+0.3 \\ -0.5}$\,K based on
pixel-by-pixel Bayesian analysis of LTE synthetic methanol spectra across MM1. 
Intriguingly, the peak CH$_{3}$OH $T_{\mathrm{ex}}=165.5\pm0.6$\,K is offset from MM1's millimetre continuum peak by $0.22\arcsec\sim880$\,{\sc au}, and a region of elevated CH$_{3}$OH $T_{\mathrm{ex}}$ coincides with free-free VLA $5.01$\,cm continuum, adding to the tentative evidence for a possible unresolved high-mass binary in MM1. In our VLA $1.21$\,cm data, we report the first NH$_{3}$(3,3) maser detections towards G19.01$-$0.03, along with candidate 25\,GHz CH$_{3}$OH $5(2,3)-5(1,4)$ maser emission; both are spatially and kinematically coincident with $44$\,GHz Class~I CH$_{3}$OH masers in the MM1 outflow. We also report the ALMA detection of candidate $278.3$\,GHz Class~I CH$_{3}$OH maser emission towards this outflow, strengthening the connection of these three maser types to MYSO outflows.
\end{abstract}




\section{Introduction}
\label{sec:intro}

High-mass stars (M$_{\ast}>8$\,M$_{\odot}$) are influential in the dynamical and chemical evolution of the interstellar medium (ISM), through their strong feedback, outflows and jets, and through enrichment of the ISM with heavy elements \citep[e.g.][]{peters17,rosen20,mignon-risse21,grudic22}.
Constraining exactly how high-mass stars form however is hampered by their natal molecular clouds being significantly more distant ($d>1$\,kpc) and more clustered ($n_{\ast}>100$\,pc$^{-3}$) than those of their low-mass counterparts (M$_{\ast}<8$\,M$_{\odot}$). Furthermore, the short pre-main sequence lifetimes \citep[$<1$\,Myrs;][]{mottram11} of high-mass stars ensures that the entirety of their formation is obscured in regions of high extinction \citep[e.g.][]{chevance20,kim21}. Now with the advent of facilities capable of high-angular resolution observations in the (sub)millimetre such as the Atacama Large Millimetre/(sub)millimetre Array (ALMA), we have the ability to resolve and disentangle the thermal emission of the early stages of high-mass star formation.

The core-fed theory of massive star formation \citep{mckeetan03,tan14} describes the monolithic collapse of virialised high-mass prestellar cores that have ceased accreting material from their surroundings, and are supported against fragmentation into low-mass cores by magnetic and turbulent pressures.
In this picture, high-mass prestellar cores \citep[e.g.][]{motte07} --  starless and self-gravitating structures thought to form in infrared dark clouds \citep[IRDCs;][]{rathjack06,peretto09} -- are the earliest stage of massive star formation.
Observationally, however, very few candidates of truly quiescent high-mass prestellar cores exist \citep[e.g.][]{cyganowski14,cyganowski22,duarte-cabral14,wang14,kong17,nony18,barnes23}, suggesting they may either be very short lived \citep[e.g.][]{motte07,kauffmann13,sanhueza19}, or not exist at all \citep[e.g.][]{motte18}.

High-mass protostars \citep[or massive young stellar objects, MYSOs, e.g.][]{hoare05,urquhart08} are instead widely associated with active star formation signatures such as 6.7\,GHz CH$_{3}$OH, 22\,GHz H$_{2}$O, and NH$_{3}$(3,3) masers \citep[e.g.][]{pillai06,urquhart11,brogan19,jones20,towner21}. Signatures of ongoing accretion are also prevalent towards MYSOs, such as high-velocity bi-polar outflows \citep[e.g.][]{beuther02,duarte-cabral13,yang22} and less commonly circumstellar accretion discs \citep[e.g.][]{beltran14,johnston15,ilee16,cesaroni17,maud18,dewangan22}.  MYSOs are typically observed to have strong (sub)millimetre continuum, and a sub-population of MYSO(s) -- classed as Extended Green Objects \citep[EGOs;][]{cyganowski08,cyganowski09} -- exhibit extended $4.5\mu$m emission attributed to shocks in outflows driven by the central MYSO(s). Many of these EGOs only exhibit weak centimetre continuum emission \citep{cyganowski11b,towner21}. 
The clump-fed theory of massive star formation \citep{bonnell01,smith09} describes the competitive accretion of clusters of initially low-mass star progenitors. High-mass star progenitors develop through the continued accretion of material by protostellar sources that find themselves at the centre of the gravitational potential of the cluster.
This requirement for a continually accreting protocluster in the clump-fed theory is a key factor that distinguishes it from the monolithic core-fed theory, and is a crucial observable for constraining models of early massive star formation  \citep[e.g.][]{cyganowski17,issac20,law22}.

Complex organic molecules (COMs) are molecular species that contain at least one carbon atom and a total of at least 6 atoms \citep[][]{herbst09}. 
They are thought to form on dust grain surfaces during both the cold collapse and "warm-up" phases of massive star formation \citep[e.g.][]{garrod06,garrod13, oberg16,garrod22}, with COMs forming earlier and at lower temperatures in models including non-diffusive chemistry \citep{garrod22}.
In general, in gas-grain astrochemical models radiative heating from MYSOs ultimately heats the grains enough to sublimate their ice mantles, releasing both COMs and simpler molecules into the gas phase where they are observable via (sub)millimetre wavelength spectral line emission \citep[e.g.][and references therein]{oberg16,garrod22}.  While COMs can also be produced by gas-phase mechanisms, production on grains dominates in recent models \citep[e.g.][]{garrod22}. 
MYSOs that are characterised by a forest of molecular lines in the (sub)millimetre are classed as hot cores \citep[e.g.][]{sanchez-monge17,sewilo18,liu21}, and typically have temperatures that exceed 100\,K \citep[e.g.][]{oberg16}.
Later in the evolution of the system (though not necessarily independent of the hot core stage), hypercompact (HC) H{\sc ii} regions form, signposted by  centimetre continuum due to the ionization of the surrounding material by the high-mass protostar(s) \citep[e.g.][]{kurtz05_HC,yang19,yang21_HC}.
With the exquisite sensitivities achievable by ALMA in the (sub)millimetre and the Karl G. Jansky Very Large Array (VLA) in the centimetre, combined studies of both the chemical and physical properties of massive star progenitors and their environments are now possible.

\begin{figure}
\centering
\includegraphics[scale=.31]{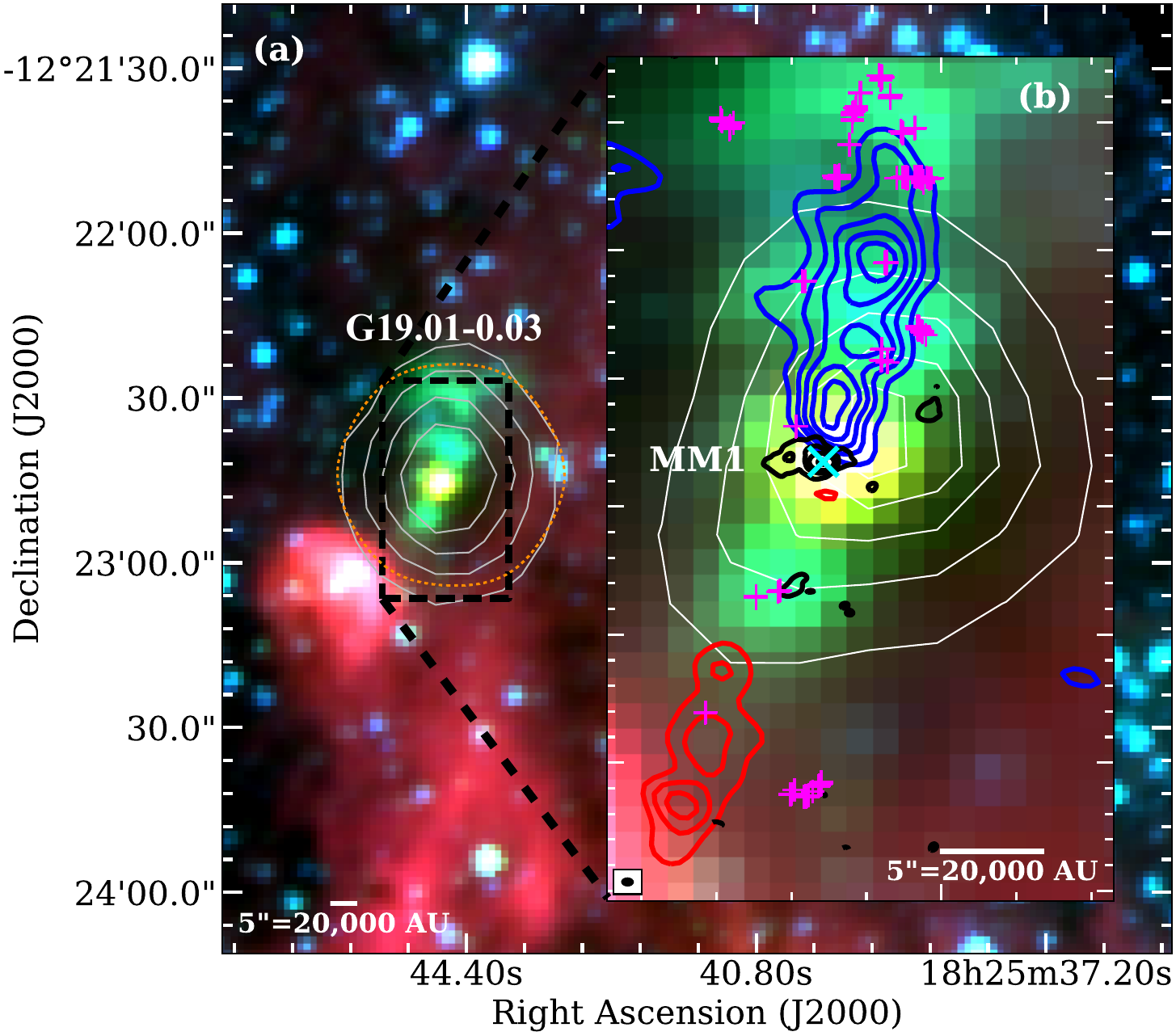}
\caption{\emph{Spitzer} GLIMPSE three-colour image (RGB: 8.0, 4.5, 3.6$\mu$m). (a) The dotted orange contour marks the 30 per cent response level of the ALMA mosaic, designed to encompass the full extent of the ATLASGAL clump. ATLASGAL 870$\mu$m emission contours are shown in grey \protect\citep[at 12, 16, 20 and 24$\sigma$, where $\sigma=0.08$\,Jy\,beam$^{-1}$;][18$\arcsec$ resolution]{schuller09}. The zoomed inset of (b) is shown by the dashed black box. (b) High-velocity blue- and red-shifted SMA $^{12}$CO(2--1) emission contours are shown in blue (7.2, 12.0, 15.6, 19.2, 22.8 Jy\,beam$^{-1}$\,km\,s$^{-1}$) and red (4.8, 7.2, 9.6 Jy\,beam$^{-1}$\,km\,s$^{-1}$) respectively \protect\citep{cyganowski11a}, and ALMA 1.05\,mm continuum contours are shown in black \protect\citep[0.00125, 0.004, 0.016, 0.200 Jy\,beam$^{-1}$;][]{williams22}.  Magenta $+$'s mark the VLA positions of 44\,GHz Class~I CH$_{3}$OH masers \protect\citep{cyganowski09}, and the cyan $\times$ marks the intensity-weighted VLA 6.7\,GHz Class~II CH$_{3}$OH maser position from \protect\citet{williams22}.  Herschel PACS $70\mu$m emission \citep{poglitsch10} of the HI-GAL clump \citep[][]{elia17} is plotted with white contours
(2.5, 4.5, 6.5, 8.5, 10.5 Jy\,pixel$^{-1}$). The PACS $70\mu$m beam is $6\arcsec\times12\arcsec$ \citep[FWHM $\sim$\,8$\arcsec$;][]{pacs_supplement}. The ALMA beam is plotted in the bottom left.}
\label{fig:g19}
\end{figure}

\subsection{Target: EGO G19.01--0.03}

The EGO G19.01--0.03 (hereafter G19.01) is an intriguing example of early massive star formation.
As in \citet{williams22}, we adopt D=$4.0\pm0.3$\,kpc, the near kinematic distance estimated using the Galactic rotation curve parameters of \citet{reid14} and the NH$_{3}$ LSRK velocity from \citet{cyganowski13}.
As shown in Figure~\ref{fig:g19}, emission from the surrounding clump is detected in the Hi-GAL and ATLASGAL surveys; the clump is also detected at $1.1$\,mm by the Bolocam Galactic Plane Survey \citep[BGPS;][]{rosolowsky10}.  
The clump mass is $\sim$1000\,M$_{\odot}$:  
the Hi-GAL catalogue of \citet{elia17} reports a clump mass, radius and temperature of 974\,M$_{\odot}$, 0.16\,pc and 17.9\,K respectively for D=3.658 kpc ($\sim$1165\,M$_{\odot}$ and 0.18\,pc scaled to D=4.0 kpc), while \citet{schuller09} report a clump mass of 1070\,M$_{\odot}$ for D=4.3 kpc based on the $870\mu$m ATLASGAL data ($\sim$926\,M$_{\odot}$ scaled to D=4.0 kpc). 
Towards the clump, a single millimetre continuum core (hereafter called MM1) is seen in isolation with the Submillimeter Array (SMA) at 1.3\,mm and with the Combined Array for Research in Millimeter-wave Astronomy (CARMA) at 3.4\,mm  \citep[at 2.4 and 5.4$\arcsec$ angular resolution respectively;][]{cyganowski11a}. 
MM1 coincides with 6.7\,GHz Class~II CH$_{3}$OH maser emission \citep{cyganowski09} placing it as a massive source \citep[e.g.][]{urquhart13a,billington20,jones20}, and is seen to drive a highly collimated, high-velocity bi-polar outflow observed with the SMA in $^{12}$CO(2--1) and with CARMA in HCO$^{+}$(1--0) and SiO(2--1) emission \citep{cyganowski11a}. 44\,GHz Class~I CH$_{3}$OH maser emission is seen to trace the edges of the outflow lobes \citep{cyganowski09}. 
At the sensitivity of the SMA and CARMA data, MM1 exhibited some hot core emission but lacked chemical richness, conspicuously lacking in emission from Oxygen-bearing COMs. Only two COMs were detected (CH$_{3}$OH and CH$_{3}$CN) with only two lines with excitation temperature $>$100\,K \citep{cyganowski11a}. 
MM1 was not detected in deep, arcsecond-resolution VLA observations at 3.6 and 1.3\,cm to $4\sigma$ limits of 0.12 and 1.04\,mJy\,beam$^{-1}$ respectively \citep{cyganowski11b}, suggesting a very low ionising luminosity capable of producing only a very small H{\sc ii} region. 
Put together, MM1 appeared as an isolated, high-mass millimetre continuum source, in a state of ongoing accretion, without strong centimetre continuum or rich hot-core line emission. As such, MM1 was until recently considered an excellent candidate for a very early stage of evolution, with potential to shed light on the core-fed theory of high-mass star formation.

The first paper of our ALMA Cycle 2 follow-up study \citep[][hereafter \paperone{}]{williams22} presented the highest angular resolution observations of G19.01 to date, at $\sim0.4\arcsec$ angular resolution in Band 7 at 1.05\,mm. With ALMA, MM1 was observed to exhibit a rich millimetre spectrum with a variety of COMs, in contrast to the earlier lower resolution and sensitivity SMA observations. 
Kinematic analysis of the strongest, most isolated ALMA-detected molecules revealed the first direct evidence of a rotationally supported accretion disc around MM1 traced by a velocity gradient perpendicular to the bi-polar outflow direction, with an enclosed mass of $40-70$\,M$_{\odot}$ within a 2000\,AU radius. 
In conjunction with new VLA observations at 5.01 and 1.21\,cm, the centimetre-millimetre spectral energy distribution (SED) was best described by a two-component model, with millimetre emission dominated by thermal dust, and the $\sim5$\,cm continuum dominated by free-free emission interpreted as a hypercompact H{\sc ii} region, placing MM1 in a later stage of evolution than that concluded with previous observations.
Furthermore, the ALMA 1.05\,mm continuum revealed for the first time the detection of four neighbouring millimetre sources in the vicinity of MM1, hinting at the possibility of the early stages of protocluster formation.

In this paper (Paper II), we use our ALMA Cycle 2 1.05\,mm and VLA 1.21\,cm observations to study the chemistry and protocluster environment of G19.01--0.03 MM1 and the properties of the newly-detected millimetre sources.
In Section~\ref{sec:observations} we describe the observations, in Section~\ref{sec:results} we present the ALMA continuum and molecular line emission, as well as VLA ammonia and methanol emission. In Section~\ref{sec:discussion} we present our modelling of the COM emission, rotation diagram analysis, continuum properties of the millimetre neighbours, and discuss MM1's chemistry in the context of sources from the literature. We summarise our main conclusions in Section~\ref{sec:conclusions}.

\section{Observations}
\label{sec:observations}

\subsection{Atacama Large Millimetre/submillimetre Array (ALMA)}
\label{sec:alma_obs}

Our ALMA Cycle 2 observations were designed to search for low-mass cores within the clump-scale gas reservoir associated with G19.01--0.03 MM1: the extent of the ALMA mosaic ($\sim$40$\arcsec\approx$0.78\,pc at D=4\,kpc) is shown in Figure~\ref{fig:g19} and observing parameters are summarised in Table~\ref{tab:obs} and below.  These observations are also described in detail in \paperone{}.  

For our observations, the ALMA correlator was configured to cover seven spectral windows (spws), including five narrow spws targeting particular spectral lines and two wide spws.  Details of the narrow spws are given in Table~\ref{tab:obs_narrowline}.  The wide spws, with central frequencies of $\sim$278.2\,GHz and $\sim$292.0\,GHz, each have a bandwidth of 1.875\,GHz, a Hanning-smoothed spectral resolution of 1.13\,MHz (1.156$\times$ the channel spacing of 0.977\,MHz because of online channel averaging in the ALMA correlator).

\begin{table}
	\small
	\centering
	\caption{Observing parameters for ALMA 1.05\,mm and VLA 1.21\,cm data.}
	\label{tab:obs}
	\begin{tabular}{lcc}
	\hline\hline
	Parameter 					    & ALMA 1.05\,mm  				& VLA 1.21\,cm \\ \hline
	Observing date 				    & 14 May 2015 				& 11-12 Nov 2013 							\\
    Project code (PI) &  2013.1.00812.S  & 13B-359  \\
                     & (C.\ Cyganowski) & (T.\ Hunter) \\
    Gain calibrator 			    & J1733--1304 				& J1832--1035						\\
	Bandpass calibrator 		    & J1733--1304 				& J1924--2914						\\
	Flux calibrator 			    & Titan$^{a}$ 					& J1331+3030							\\
	On-source integration time 			    & 44\,min 			& 139\,min 		    		\\
	Number of antennas 			    & 37 						& 25 											\\
	Antenna configuration 		    & C43-3/(4)					& B 						 						\\
	Phase Centre (J2000): 			&  		                    &                       						\\
	\,\,\,\,\,\,R.A. ($^{\mathrm{h\,m\,s}}$)  & 18:25:44.61$^{b}$               & 18:25:44.80                \\
	\,\,\,\,\,\,Dec. ($^{\circ}$ $'$ $''$)    & -12:22:44.00$^{b}$              & -12:22:46.00             \\
	Projected baseline lengths 	    & 20--533\,m 				& 0.14--9.98\,km 			 			\\
		    & 19--508\,$k\lambda$  & 12--825\,$k\lambda$   \\
	Mean frequency$^c$ 		 		    & 285.12\,GHz 				& 24.81\,GHz 							\\
	Mean wavelength$^c$	 		 		    & 1.05\,mm		 			& 1.21\,cm 						\\
    Number of pointings & 7 & 1\\
    Field of view$^{c,d}$ & $\sim$40$\arcsec$ & 1.8$'$ 						\\
	Synthesised beam$^{c}$ 			    & $0\farcs52 \times 0\farcs35$ 	& $0\farcs33 \times 0\farcs22$	\\
	Beam position angle$^{c,e}$  		    & 88.4$^\circ$ 				& 0.5$^\circ$			 			\\
	Maximum Recoverable Scale$^{f}$ & $4\farcs2$ 					& $4\farcs5$ 								\\
	Continuum rms noise$^{g}$ 	    & 0.25\,mJy\,beam$^{-1}$ 	& 6.0\,$\mu$Jy\,beam$^{-1}$	 \\
	
	\hline
	\end{tabular}
	\begin{flushleft}
 		$^a$ Using Butler-JPL-Horizons 2012 models.\\
	    $^b$ For the central pointing of the mosaic.\\
	    $^c$ For the continuum image.\\
	    $^d$ ALMA: to 30\% level of mosaic response.  VLA: primary beam FWHP at mean frequency.\\
	    $^e$ Measured East of North i.e. positive in the anti-clockwise direction.\\
		$^f$ Calculated from the fifth percentile shortest baseline (as stated in the ALMA Technical Handbook) and mean frequency, using \texttt{au.estimateMRS} from the analysisUtils Python package.\\
		$^g$ Estimated from emission-free regions within the 30\% response level of the ALMA mosaic. 
	\end{flushleft}
\end{table}

As detailed in \paperone{}, the data were calibrated using the \textsc{casa} 4.2.2 version of the ALMA calibration pipeline and line-free channels were identified using the approach of \citet{brogan16,cyganowski17}.  These line-free channels were used to construct a pseudo-continuum dataset and to perform continuum subtraction in the \emph{u,v}-plane.  For the narrow spw targeting C$^{33}$S, this process was problematic due to wide lines and possible absorption \citep[see also \paperone{} and][]{cyganowski17} and we excluded this spw -- which overlaps one of the wide spws in our tuning -- from the pseudo-continuum dataset and our line analysis.  The aggregate continuum bandwidth of the final pseudo-continuum dataset is $\sim$1.6\,GHz.  The continuum data were iteratively self-calibrated and were imaged using Briggs weighting with a robust parameter of 0 and multi-frequency synthesis; the key parameters of the resulting image are listed in Table~\ref{tab:obs}.  The synthesised beamsize of the continuum image (0\farcs52$\times$0\farcs35) corresponds to a physical scale of 2080$\times$1400\,AU at 4\,kpc.  
As our observations included only the ALMA 12\,m array, the maximum recoverable scale is $4.2\arcsec{\sim}16,800$\,{\sc au} (at 4\,kpc).

\begin{table*}
	\small
	\centering
	\caption{Observing and image cube parameters for narrow spectral windows}
	\label{tab:obs_narrowline}
    \setlength\tabcolsep{4.75pt}
	\begin{tabular}{lccccccccc}
	\hline\hline
    Telescope & Targeted line & Line Rest Frequency$^{a}$ & E$_{u}/k_{B}$$^{a}$ & Bandwidth & $\Delta\nu$$^{b}$ & $\Delta$v$^{c}$ & Synthesised beam & \multicolumn{2}{c}{rms noise$^{d}$} \\
     & & (GHz) & (K) & (MHz) & (kHz) & (km s$^{-1}$) & $\arcsec\times\arcsec$ [$^{\circ}$] & (mJy beam$^{-1}$) & (K) \\
    \hline
    ALMA & N$_{2}$H$^{+}$(3--2) & 279.5117491 & 26.8 & 468.75 & 122.07 & 0.5 & 0.57 $\times$ 0.41 [83.2] & 6.1 & 0.41 \\
    ALMA & DCN (4--3) & 289.644917 & 34.8 & 117.1875 & 122.07 & 0.5 & 0.55 $\times$ 0.40 [82.0] & 6.3 & 0.42 \\
    ALMA & $^{34}$SO 6$_{7}$--5$_{6}$ & 290.562238 & 63.8 & 117.1875 & 122.07 & 0.5 &  0.55 $\times$ 0.40 [82.2] & 6.0 & 0.39 \\
    ALMA & H$_{2}$CO $4_{0,4}-3_{0,3}$ & 290.623405 & 34.9 & 117.1875 & 122.07 & 0.5 & 0.55 $\times$ 0.40 [82.3] & 6.0 & 0.39 \\
    ALMA & C$^{33}$S (6--5) & 291.485935 & 49.0 & 117.1875 & 122.07 & 0.5 & 0.55 $\times$ 0.40 [82.4] & ... & ... \\
    VLA & NH$_3$ (1,1) & 23.694496 & 24.4 & 8.0 & 15.625  & 0.4 & 0.56 $\times$ 0.54 [28.0] &  1.24 & 8.93 \\ 
    VLA &  NH$_3$ (2,2) &  23.722631 & 65.6 & 8.0 & 15.625  & 0.4 & 0.56 $\times$ 0.54 [28.5] & 1.18 & 8.47\\
    VLA & NH$_3$ (3,3) & 23.870130 & 124.7 &  8.0 & 15.625  & 0.4 & 0.56 $\times$ 0.54 [30.5] & 1.13 & 8.01\\ 
    VLA &  NH$_3$ (5,5) &  24.532985 & 296.5 &  8.0 & 15.625  & 0.4 & 0.55 $\times$ 0.54 [38.4] & 1.10 & 7.52 \\ 
    VLA &  NH$_3$ (6,6) & 25.056025 & 409.2 & 4.0 & 15.625  & 0.4 & 0.55 $\times$ 0.54 [41.1] & 0.95 & 6.23 \\ 
    VLA &  NH$_3$ (7,7) &  25.715182 & 539.7 & 4.0 & 15.625 & 0.4 & 0.54 $\times$ 0.53 [47.8] &  1.06 & 6.84 \\ 
    VLA$^{e}$ & CH$_3$OH 3(2,1)-3(1,2) & 24.928707 & 36.2 & 2.0 &  15.625 & 0.4 &  0.56 $\times$ 0.55 [31.5] & 1.04 & 6.76\\ 
    VLA$^{e}$ & CH$_3$OH 5(2,3)-5(1,4) & 24.9590789 & 57.1 &  2.0 &  15.625 & 0.4 & 0.55 $\times$ 0.54 [49.7] & 1.01 & 6.67\\ 
    VLA$^{e}$ & CH$_3$OH 8(2,6)-8(1,7) & 25.2944165  & 105.8 & 2.0 &  15.625 & 0.4 & 0.57 $\times$ 0.55 [52.8] & 0.85 & 5.18 \\
    VLA$^{e}$ & CH$_3$OH 10(2,8)-10(1,9) & 25.8782661 & 150.0 &  2.0 &  15.625 & 0.4 & 0.56 $\times$ 0.54 [65.0] & 1.19 & 7.18 \\
    \hline
    \end{tabular}
	\begin{flushleft}
  		$^a$ From CDMS \citep{cdms,cdms2} for lines observed with ALMA and from the TopModel line list for NH$_3$ lines, both accessed via the NRAO spectral line catalogue (Splatalogue; \url{https://splatalogue.online/}).  CH$_3$OH line frequencies are from \citet{muller04}.\\
        $^b$ Channel spacing.  For the ALMA data, the Hanning-smoothed spectral resolution is 0.244 MHz for all of the narrow spws.  \\ 
        $^c$ Velocity channel width of image cubes; see \S\ref{sec:alma_obs} and \S\ref{sec:vla_obs}.\\
        $^d$ Median rms noise estimated from emission-free channels; the rms noise is up to $\sim$1.5 times higher in channels with bright and/or complex emission.  The conversion to brightness temperature assumes the Rayleigh-Jeans approximation: \begin{math} T=1.222\times 10^3 \frac{I}{\nu^2\theta_{\mathrm{maj}}\theta_{\mathrm{min}}} \end{math}, where $I$ is the rms in mJy beam$^{-1}$, $\nu$ is the frequency in GHz, and $\theta_{\mathrm{maj}}\times\theta_{\mathrm{min}}$ is the synthesised beam size in arcseconds.\\
        $^e$ As in \cite{towner17}, these 25\,GHz CH$_{3}$OH lines will be referred to in the main text by the following shorthand notation, given by the first two values of the upper state quantum number of each transition respectively: 3$_{2}$, 5$_{2}$, 8$_{2}$, 10$_{2}$. 
 	  \end{flushleft}
\end{table*}

After applying the solutions from the continuum self-calibration, the continuum-subtracted line data were imaged with Briggs weighting with a robust parameter of 0.5.  For the narrow spws, which were imaged with 0.5 km s$^{-1}$ channels for better sensitivity to faint emission, the synthesised beamsizes and rms noise levels of the image cubes are listed in Table~\ref{tab:obs_narrowline}.  For the wide spws, which were imaged with the native channel spacing, the synthesised beamsizes are 0\farcs60$\times$0\farcs43 [P.A. 79.8$^\circ$] and 0\farcs57$\times$0\farcs41 [P.A. 78.9$^\circ$] for the spws centred at $\sim$278.2\,GHz and $\sim$292.0\,GHz, respectively, and the rms noise is $\sim$3 mJy beam$^{-1}$ in emission-free channels (in channels with complex emission, the rms is up to $\sim$1.5 times higher; see also \paperone{}).  All measurements were made from images corrected for the response of the primary beam.

\subsection{Karl G. Jansky Very Large Array (VLA)}
\label{sec:vla_obs}

In this paper, we present our K-band VLA spectral line observations of G19.01--0.03.  Our VLA tuning included 10 narrow spws targeting NH$_3$ and CH$_3$OH lines observed in other EGOs, including lines that exhibit maser activity in some EGOs \citep[e.g.][see also \S\ref{sec:vla_ammonia}; \S\ref{sec:vla_methanol}]{brogan11,towner17}.  The VLA observing parameters are summarised in Table~\ref{tab:obs} and details of the narrow spws are given in Table~\ref{tab:obs_narrowline}.  The VLA K-band tuning also included 16$\times$0.128\,GHz spws for continuum, as detailed in \paperone{} which presented the continuum results; for completeness, key parameters of the VLA 1.21\,cm continuum image are included in Table~\ref{tab:obs}.  

As explained in \paperone{}, the VLA data were calibrated using the \textsc{casa} 4.7.1 version of the VLA calibration pipeline.  The data were Hanning smoothed, and phase-only self-calibration was performed using the channel with the strongest NH$_3$(3,3) emission (after continuum subtraction in the \emph{u,v}-plane).  
These solutions were then applied to all of the line data (as well as to the continuum; see \paperone{}).
For each narrowband spw, continuum subtraction was performed in the \emph{u,v}-plane and the continuum-subtracted line data were imaged with 0.4 km s$^{-1}$ channels, Briggs weighting with a robust parameter of 0.5, and a \emph{uv} taper of 200 k$\lambda$ to improve the brightness temperature sensitivity. 
The synthesised beamsizes and rms noise levels of the resulting image cubes are presented in Table~\ref{tab:obs_narrowline}.
Measurements were made from images corrected for the primary beam response.

\section{Results}
\label{sec:results}

\subsection{ALMA 1.05 mm continuum emission}

\begin{figure}
\centering
\includegraphics[scale=.73]{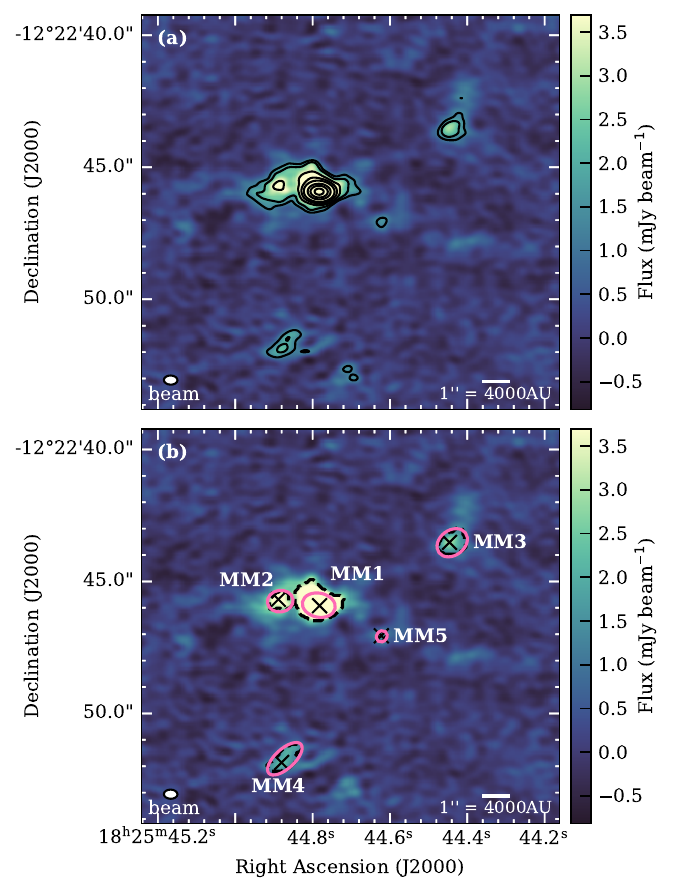}
\caption{(a) ALMA 1.05\,mm continuum image, corrected for the primary beam response, as shown in \paperone{}. The field shown is a sub-region of that mosaiced, which contains all emission detected $\geq5\sigma$. Black contours are plotted at [5, 8, 16, 32, 64, 200, 400 and 800]$\times\sigma$, where $\sigma=0.25$\,mJy\,beam$^{-1}$. The synthesised beam and scalebar are plotted in the bottom-left and bottom-right respectively. (b) Same as (a), but with the outlines of the extracted dendrogram structures in dashed black, and ellipses representing the source sizes (Table~\ref{tab:leaves}) in pink. The peak positions of the sources (Table~\ref{tab:leaves}) are marked by the black $\times$s.}         
\label{fig:continuum}
\end{figure}

In \paperone{}, we presented the ALMA 1.05\,mm continuum towards MM1, and noted the detection of a further four continuum sources in the field for the first time (see Figure~\ref{fig:continuum}), named MM2...MM5 in order of decreasing peak intensity. Their observed properties (as well as the observed properties of MM1 as presented in \paperone{}) are listed in Table~\ref{tab:leaves}, with their FWHM extents represented by pink ellipses in Figure~\ref{fig:continuum}b.  As detailed in \paperone{}, sources were extracted using the {\sc astrodendro} algorithm \citep{rosolowsky08dendro}, with a minimum isocontour value ($I_{\mathrm{min}}$) of $5\sigma_{\mathrm{rms}}$ (where $\sigma_{\mathrm{rms}}=0.25$\,mJy\,beam$^{-1}$), minimum isocontour spacing ($\Delta I_{\mathrm{min}}$) of $1\sigma_{\mathrm{rms}}$, and minimum size of a structure ($n_{\mathrm{pix}}$) of $\approx n_{\mathrm{pix,beam}}/2$ (i.e. half the beam size, where $n_{\mathrm{pix,beam}}\approx50$). Two sources to the south-west of MM4 are detected with peak emission $>5\sigma_{\mathrm{rms}}$ (Figure~\ref{fig:continuum}) but are only extracted by the dendrogram algorithm if the parameters are dropped to $\Delta I_{\mathrm{min}} = 0.9\sigma_{\mathrm{rms}}$ and $n_{\mathrm{pix}} = 15$ pixels, meaning that they are only equivalent to a third of a beam in size. We therefore do not consider these firm detections. MM2...MM5 have angular separations from MM1 of 1.6, 5.5, 6.1 and 2.6$\arcsec$ respectively, ranging between $0.03-0.12$\,pc at the 4\,kpc distance, marking the first detection of other millimetre sources within the parent clump of MM1. 
MM4 appears non-gaussian in its emission morphology, whilst MM2 lies within a common contour to MM1 with lower surface-brightness emission connecting the two sources. This suggests that MM2 may be fragmenting out of material that also feeds MM1.

\begin{table*}
    \centering
    \caption{Observed properties of extracted 1.05\,mm continuum sources.}
    \begin{threeparttable}
    \label{tab:leaves}
    \setlength\tabcolsep{12.4pt}
    \begin{tabular}{ccccccc}
    \hline\hline
     Source 	     & \multicolumn{2}{c}{J2000.0 Coordinates$^a$}	    & Peak		            & Integ.            &  Source size $^{c}$  		        & Source size$^{c}$ 			    \\ 
        	             &  $\alpha$  	& $\delta$ 	                & intensity $^{b}$	    & flux $^{b}$  		& Maj. $\times$ Min. [P.A.]		           &               \\
     	         	             &  ($^{\mathrm{h\,m\,s}}$)     & ($^{\circ}$ $'$ $''$)                   & (mJy\,beam$^{-1}$)    & (mJy)             & (arcsec~$\times$~arcsec [$^{\circ}$])                            & ({\sc au})                           \\ \hline
    MM1       & 18:25:44.782	& -12:22:45.92 				& $266.3$ 				& $303.1$			& $1.15 \times 0.84~[78.7] $ 		        & $4600\times3360$ 		           	\\
    MM2       & 18:25:44.888	    & -12:22:45.68 				& $4.7$		& $8.1$	    & $0.83 \times 0.74~[-82.1] $ 		    & $3310 \times 2970$ 		                    	    \\     
    MM3       & 18:25:44.446	    & -12:22:43.52 		        & $3.0$		& $7.1$     & $1.14 \times 0.88 ~[-53.0]$	                & $4560\times3530$				 \\ 
    MM4       & 18:25:44.880	    & -12:22:51.86 		        & $2.3$		& $6.0$     & $1.56 \times 0.65 ~[-47.7]$	                & $6200\times2610$				 \\
    MM5       & 18:25:44.622     & -12:22:47.06 		        & $1.6$		& $-$     & $-$	                & $-$				 \\ \hline
    \end{tabular}
    \begin{tablenotes}
        \item[$a$] Peak position. The number of significant figures reflects a one pixel uncertainty.
        \item[$b$] Evaluated within the intensity-weighted second moment size (not the total dendrogram structure). For sources smaller than a beam, integrated fluxes are marked ``$-$''.
    	\item[$c$] Deconvolved major and minor axes sizes; position angle is measured East of North i.e. positive in the anti-clockwise direction.  Sizes are the intensity-weighted second moment, converted to FWHM (see Paper {\sc i}). Sources smaller than a beam are marked ``$-$''.
    \end{tablenotes}
    \end{threeparttable}
\end{table*}

\subsection{Line emission}

\subsubsection{COMs towards MM1 with ALMA}
\label{sec:lines}

\begin{figure*}
\centering
\includegraphics[scale=.52]{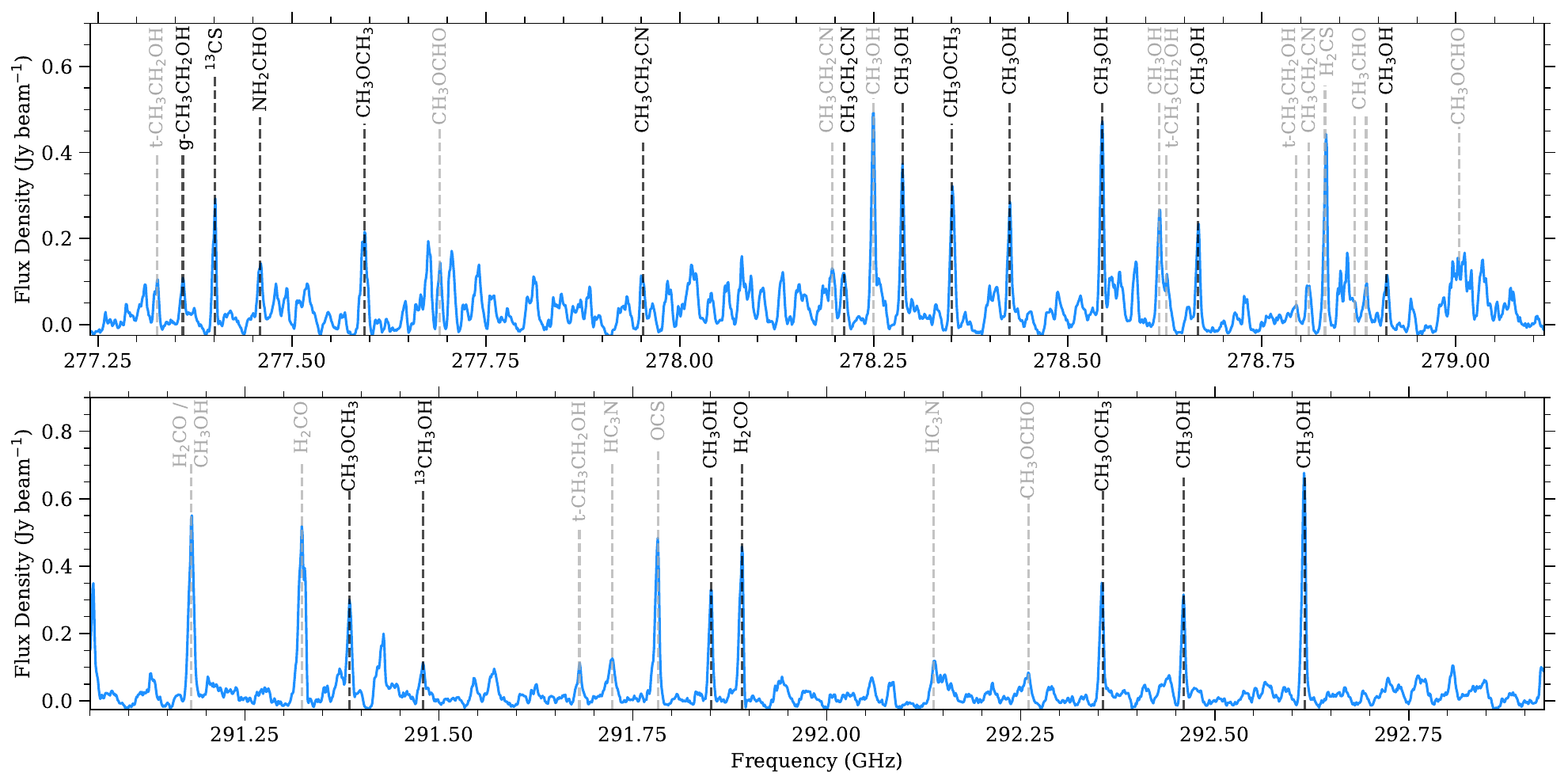} 
\caption{ALMA spectra towards the 1.05\,mm continuum peak of MM1 in the broad spectral windows centred at 278.209\,GHz and 292.021\,GHz. Unlabelled lines could not be confidently identified, mostly due to blending at the coarse spectral resolution of our observations. Lines labelled in black were used for the kinematic analysis presented in \paperone{}, while others are labelled in grey. All marked lines are listed in Table~\ref{tab:lines}.}       
\label{fig:spec6}
\end{figure*}

\begin{table*}
	\footnotesize
	\centering
	\caption{Properties of spectral lines identified in the broad ALMA spectral windows towards MM1, arranged by decreasing $E_{u}$/$k_{B}$.}
	\label{tab:lines}
    \setlength\tabcolsep{12.4pt}
	\begin{tabular}{llcccccc}
	\hline\hline
	Species$^{a}$ 						    & Transition  					    & Frequency	        & $E_{u}$/$k_{B}$ & $S_{ij}\mu^{2}$ & $g_{u}$         & Catalogue$^{b}$ & 	\paperone{}?$^{c}$  	\\
	                                    &                                   & (GHz)             & (K)          &  (D$^2$)          &      &                   &                    \\ \hline
	\textbf{CH$_{3}$OH (v$_{t}$ = 0)}			& 23$_{4,19}$--22$_{5,18}$		    & 278.96513         & 736.0        & 7.05321            &  47               & JPL               & Y         \\ 
	\textbf{CH$_{3}$OH (v$_{t}$ = 0)}			& 21$_{-2,20}$--20$_{-3,18}$	  	& 278.47989			& 563.2        & 6.45393              & 43              & JPL 			    & Y			\\
	HC$_{3}$N (v7=1)				    & J = 32--31, I = 1f			  	& 292.19837			& 551.4	       & 443.01616                  &  65                 & CDMS			    & N			\\
	HC$_{3}$N (v7=1)				    & J = 32--31, I = 1e			  	& 291.78201			& 551.1	       & 442.99035                      & 65              & CDMS			    & N			\\
	CH$_{3}$OH (v$_{t}$ = 0)	        & 17$_{6,12}$--18$_{4,13}$-\,-	    & 291.90814 		& 548.6		   & 3.98258                   & 35                 & JPL 			    & Y			\\ 
	\textbf{CH$_{3}$OH (v$_{t}$ = 0)}			& 18$_{5,13}$--19$_{4,16}$-\,-	    & 278.72314 		& 534.6	       & 5.26251               & 37              & JPL 			    & Y			\\
	\textbf{CH$_{3}$OH (v$_{t}$ = 0)}			& 18$_{5,14}$--19$_{4,15}$+\,+	    & 278.67303 		& 534.6	       & 5.26201               & 37              & JPL 			    & N			\\
	CH$_{3}$OH (v$_{t}$ = 1)			& 10$_{1,10}$--9$_{0,9}$ 		    & 292.51744 		& 418.8        & 5.01720                  & 21                  & JPL 			    & Y			\\
	CH$_{3}$OCHO (v = 1)                & 23$_{4,19}$--22$_{4,18}$A         & 292.31711         & 365.4        & 59.03484              & 94              & JPL               & N         \\
	t-CH$_{3}$CH$_{2}$OH                & 27$_{6,22}$--27$_{5,23}$          & 278.84892         & 363.7        & 28.22445             & 55              & JPL               & N         \\
	\textbf{CH$_{3}$OH (v$_{t}$ = 0)}			& 14$_{4,10}$--15$_{3,12}$		    & 278.59908			& 339.6	       & 4.14979              &  29              & JPL			    & Y			\\
	CH$_{3}$OH (v$_{t}$ = 0)			& 15$_{1,0}$--14$_{2,0}$		    & 291.24057			& 295.3	       & 5.70014              &  31                 & JPL			    & N			\\
	t-CH$_{3}$CH$_{2}$OH                & 23$_{6,17}$--23$_{5,18}$          & 278.68234         & 277.5        & 23.74792           & 47                  & JPL               & N         \\
	CH$_{3}$CH$_{2}$CN (v = 0)          & 31$_{7,24}$--30$_{7,23}$		    & 278.00758			& 267.8		   & 435.64965          & 63                 & JPL			    & Y			\\ 
	CH$_{3}$CH$_{2}$CN (v = 0)		    & 31$_{6,25}$--30$_{6,24}$		    & 278.26670			& 253.4		   & 441.91785          & 63                  & JPL			    & Y			\\
	CH$_{3}$CH$_{2}$CN (v = 0)		    & 31$_{6,26}$--30$_{6,25}$		    & 278.25123			& 253.4		   & 441.86236          & 63                   & JPL			    & N			\\
	CH$_{3}$CH$_{2}$CN (v = 0)		    & 31$_{5,26}$--30$_{5,25}$		    & 278.86581			& 241.4		   & 447.13304          & 63                 & JPL			    & N			\\
	g-CH$_{3}$CH$_{2}$OH    		    & 16$_{6,10}$--15$_{6,9}$ (v$_{t}$=0-0)  & 277.41431	& 213.8		   & 21.96640          & 33                          & JPL			    & Y			\\ 
	g-CH$_{3}$CH$_{2}$OH		        & 16$_{4,12}$--15$_{4,11}$ (v$_{t}$=0-0) & 278.64299	& 189.7		   & 23.96604          & 33                          & JPL			    & Y			\\
	CH$_{3}$OCHO (v = 0)                & 24$_{2,22}$--23$_{2,21}$E         & 279.05752         & 178.5        & 62.00988     &  98                              & JPL               & N         \\
	CH$_{3}$OCHO (v = 0)                & 24$_{2,22}$--23$_{2,21}$A         & 279.06596         & 178.5        & 62.01870     &  98                              & JPL               & N         \\
	OCS (v = 0)						    & J = 24--23					  	& 291.83965			& 175.1	       & 12.27714     &  49                           & CDMS			    & N			\\
	CH$_{3}$OCHO (v = 0)                & 22$_{6,16}$--21$_{6,15}$E         & 279.05067         & 175.0        & 54.25478     &  90                          & JPL               & N         \\
	CH$_{3}$OCHO (v = 0)                & 22$_{6,16}$--21$_{6,15}$A         & 279.07471         & 175.0        & 54.26998     &  90                & JPL               & N         \\
	CH$_{3}$OCHO (v = 0)                & 23$_{4,20}$--22$_{4,19}$A         & 277.74543         & 173.4        & 58.74803     &  94                          & JPL               & N         \\
	H$_{2}$CO                           & 4$_{3,2}$--3$_{3,1}$              & 291.38049         & 140.9        & 28.54373     & 27                           & CDMS              & N         \\
	t-CH$_{3}$CH$_{2}$OH                & 17$_{1,16}$--16$_{2,15}$          & 277.38114         & 131.5        & 7.54333     & 35                           & JPL               & N         \\     
	CH$_{3}$OCH$_{3}$       			& 16$_{1,16}$--15$_{0,15}$ (EA)	    & 292.41225			& 120.3        & 70.75903     & 66                          & JPL		        & Y			\\ 
	NH$_{2}$CHO						    & 13$_{3,10}$--12$_{3,9}$		    & 277.51403			& 119.8		   & 482.76031     & 81                            & JPL			    & Y			\\
	CH$_{3}$OH (v$_{t}$ = 0)$^{d}$		& 9$_{-1,9}$--8$_{0,8}$			    & 278.30451			& 110.0        & 5.82054          & 19                         & JPL 			    & N 		\\
	CH$_{3}$CHO (v$_{t}$ = 0)           & 15$_{-1,15}$--14$_{-1,14}$ (E)	& 278.92443			& 109.8	       & 188.77851     & 62                            & JPL			    & N	 		\\
	CH$_{3}$CHO (v$_{t}$ = 0)           & 15$_{1,15}$--14$_{1,14}$ (A)	    & 278.93944			& 109.7	       & 188.63110     & 62          & JPL			    & N	 		\\
	CH$_{3}$OCH$_{3}$       			& 13$_{2,12}$--12$_{1,11}$ (EE)	    & 291.44307			& 88.0	       & 116.69078    &  216                   & JPL			    & Y	 		\\ 
	H$_{2}$CO $^{d}$				    & 4$_{2,2}$--3$_{2,1}$			    & 291.94807			& 82.1		   & 16.31005     & 9                    & CDMS			    & Y			\\
	H$_{2}$CO                           & 4$_{2,3}$--3$_{2,2}$              & 291.23777         & 82.1         & 16.30879     & 9                   & CDMS              & N         \\
	CH$_{3}$OCH$_{3}$ 			        & 12$_{2,11}$--11$_{1,10}$ (EE)	    & 278.40706			& 76.3	       & 103.86585     & 200                 & JPL			    & Y			\\
	H$_{2}$CS $^{d}$				    & 8$_{1,7}$--7$_{1,6}$	 		    & 278.88640			& 73.4		   & 64.08219     & 51                   & JPL			    & N			\\
	CH$_{3}$OH (v$_{t}$ = 0) $^{d}$	    & 6$_{1,5}$--5$_{1,4}$ -\,-		    & 292.67291			& 63.7 		   & 4.72131     & 13                   & JPL			    & Y			\\
	$^{13}$CH$_{3}$OH (v$_{t}$ = 0)     & 3$_{2,2}$--4$_{1,3}$ 			    & 291.53662 		& 51.4 		   & 0.69489     & 7                  & CDMS 			    & Y			\\
	t-CH$_{3}$CH$_{2}$OH                & 9$_{3,6}$--8$_{2,7}$              & 291.73815         & 49.2         & 7.83704     & 19                & JPL               & N         \\
	$^{13}$CS (v = 0)                   & 6--5                              & 277.45540         & 46.6         & 46.00530     & 26               & CDMS              & Y         \\         
	CH$_{3}$OCH$_{3}$                   & 7$_{3,5}$--6$_{2,4}$ (EE)         & 277.64832         & 38.1         & 65.30773     & 120               & JPL               & N         \\
	\textbf{CH$_{3}$OH (v$_{t}$ = 0)}			& 2$_{-2,1}$--3$_{-1,3}$		  	& 278.34222			& 32.9         & 0.32917         & 5   & JPL			    & Y			\\
	\hline
	\end{tabular}
    \begin{flushleft}
		$^a$ Bold typeface indicates the lines used for the analysis presented in Sections~\ref{sec:synthspec}--\ref{sec:synthspec_images}.\\
        $^b$ CDMS \citep{cdms} and JPL \citep{jpl}, accessed via the NRAO spectral line catalogue (Splatalogue; \url{https://splatalogue.online/}).\\
		$^c$ Lines used for the kinematic analysis in \paperone{} are marked with a ``Y''. Lines that are newly presented here are marked with an ``N''.\\
		$^d$ Outflow tracing, see \S\ref{sec:extended_emission}.
    \end{flushleft}
\end{table*}

A forest of molecular lines is observed towards MM1 with ALMA, as seen in the wideband spectra shown here in Figure~\ref{fig:spec6}. We follow the criteria presented by \cite{herbst09} in identifying molecular lines \citep[also see][]{maret11}.
As outlined in \paperone{}, we attribute peaks in emission to cataloged rest frequencies from the JPL \citep{jpl} and CDMS \citep{cdms} databases at the source systemic velocity \citep[$59.9\pm1.1$\,km\,s$^{-1}$;][]{cyganowski11a}. We further produced LTE synthetic spectra of the molecular emission \citep[using the Weeds extension of {\sc class;}][]{maret11} -- a molecular transition was positively identified when all lines for that species predicted in the synthetic spectrum were present in the observed spectrum, for typical model parameters expected of a hot core (e.g.\ $T>100$\,K). The use of LTE synthetic spectra also allowed the identification of some emission peaks that were consistent with multiple line rest frequencies within our spectral resolution ($\sim$1\,km\,s$^{-1}$ in the wide spws), for example blended lines and lines with shoulder features.
Using this approach, we identify 43 line transitions from 11 different species in the wide ALMA spws at $\sim$278\,GHz and $\sim$292\,GHz (see Figure~\ref{fig:spec6} and Table~\ref{tab:lines}), including isolated lines, blended lines, and lines with shoulder features.
We note that many lines remain unidentified despite their strong detection above the noise, generally due to rest frequencies and/or synthetic line profiles that cannot be confidently distinguished at our spectral resolution.
Of particular note, we report compact emission towards MM1 (see Figure 3 of \paperone{}) from a range of complex organic molecules (COMs), including Oxygen and Nitrogen-bearing COMs such as CH$_{3}$OCH$_{3}$, CH$_{3}$CHO, CH$_{3}$OH, NH$_{2}$CHO and CH$_{3}$OCHO.
We also include in our analysis of MM1's line emission (\S\ref{sec:all_mm1_ch3oh} and \S\ref{sec:abundances}) additional lines from three of these species that are serendipitously included in the narrow, targeted ALMA bands: CH$_{3}$OCHO ($27_{1,27} - 26_{0,26}$) with $\nu_{\mathrm{rest}}=289.62659$\,GHz and $E_{u}/k_{B}=385.4$\,K (blended with the CH$_{3}$OH($6_{1,2} - 5_{1,2}$) line with $\nu_{\mathrm{rest}}=289.62430$\,GHz and $E_{u}/k_{B}=731.4$\,K), CH$_{3}$OH ($11_{2,10} - 10_{3,7}$) with $\nu_{\mathrm{rest}}=279.35193$\,GHz and $E_{u}/k_{B}=190.9$\,K, and OCS ($23 -22$) with $\nu_{\mathrm{rest}}=279.6853$\,GHz and $E_{u}/k_{B}=161.1$\,K.  The majority of the lines targeted with narrow spws (\S\ref{sec:alma_obs} and Table~\ref{tab:obs_narrowline}) exhibit extended emission and are discussed in Section~\ref{sec:extended_emission}. The exception is $^{34}$SO 6$_{7}$--5$_{6}$, which exhibits compact emission but which we do not include in our analyses due to line blending in the narrowband cube.

In contrast, in the SMA data presented by \cite{cyganowski11a}, MM1 was relatively line-poor (see \S\ref{sec:intro}): the only detected COM emission in their 2\,GHz-wide bands, centred at $\sim$\,220 and $\sim$\,230\,GHz, was from CH$_{3}$OH and CH$_{3}$CN. 
With ALMA, we also detect molecular lines with up to four times higher $E_{u}/k_{B}$ than detected with the SMA (e.g.\ CH$_{3}$OH($23_{4,19}-22_{5,18}$) at E$_{u}/k_{B}=736$\,K), and identify 19 lines with E$_{u}/k_{B}>200$\,K.   
The relative dearth of molecular lines in the SMA spectra is likely attributable to a combination of sensitivity and beam dilution; our $\sim$\,0.4$\arcsec$-resolution ALMA observations improve on the 2.4$\arcsec$-resolution SMA observations by a factor of $\sim$30 in beam area, and a factor of $\sim$1.8 in brightness temperature sensitivity. 
Comparing our ALMA detections with those from the sensitive, 29$\arcsec$-resolution 1~mm single-dish survey of \citet{he12} confirms the importance of beam dilution: their observations of G19.01$-$0.03 have a 1$\sigma$ rms of $0.015-0.022$\,K (compared to 0.25\,K and 0.44\,K for the ALMA and SMA observations, respectively), but the only COM detected is CH$_3$OH. 
In sum, our ALMA results affirm the hot core classification of G19.01--0.03 MM1 and illustrate the importance of sensitive, high-resolution observations for studying the chemistry of MYSOs.

\subsubsection{Extended emission with ALMA}
\label{sec:extended_emission}

\begin{figure*}
\centering
\includegraphics[scale=.39]{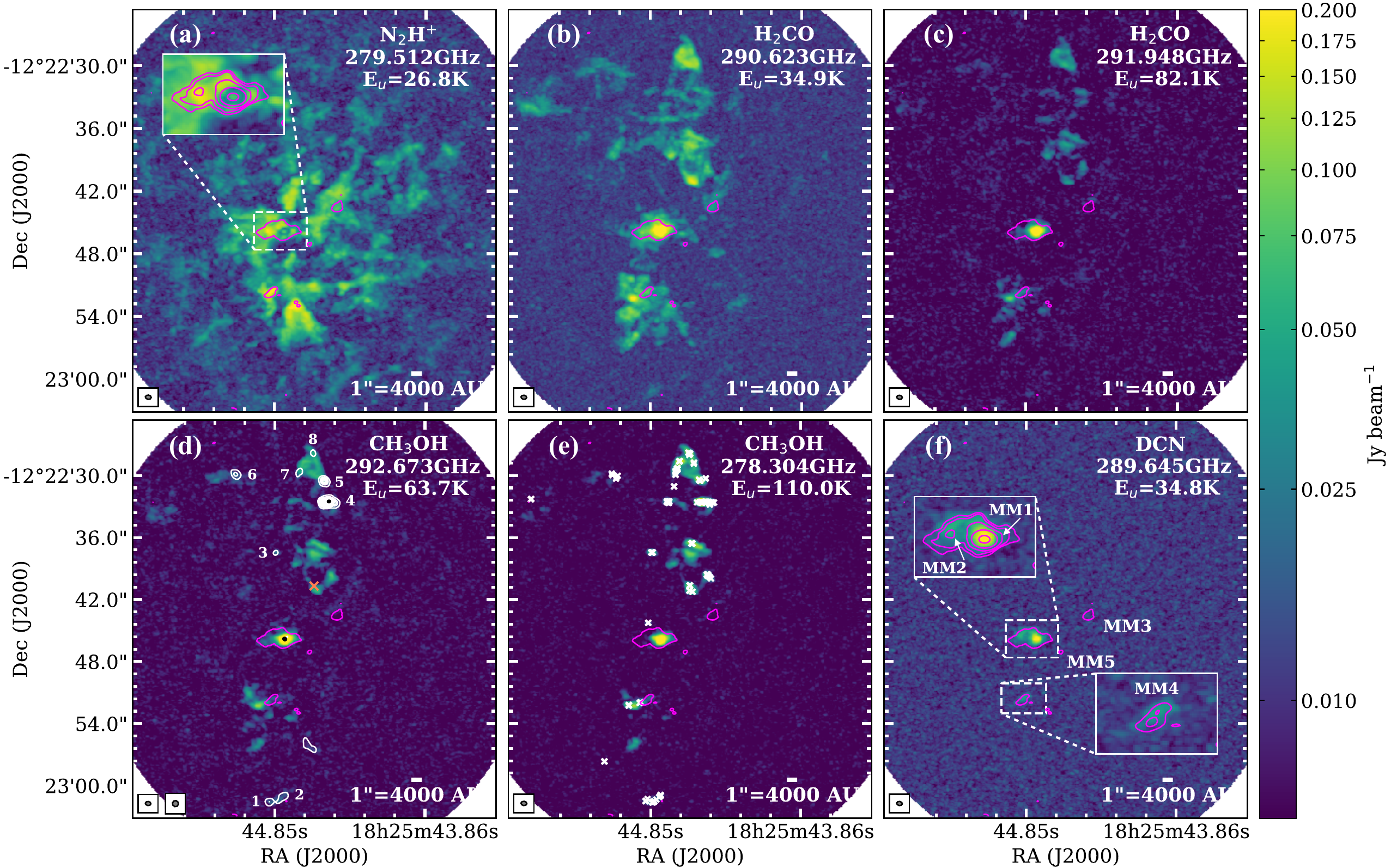}
\caption{ALMA peak intensity maps of (a) N$_{2}$H$^{+}$($3-2$), (b) H$_{2}$CO($4_{0,4}-3_{0,3}$), (c) H$_{2}$CO($4_{2,3}-3_{2,1}$), (d) CH$_{3}$OH $\mathrm{v}_{t}=0$ ($6_{1,5}-5_{1,4}$), (e) CH$_{3}$OH $\mathrm{v}_{t}=0$ ($9_{-1,9}-8_{0,8}$), and (f) DCN v=0 ($4-3$), in units of Jy\,beam$^{-1}$ shown on a logarithmic scale (images shown are not primary beam corrected). Millimetre sources are labelled in (f), where two zoomed insets are also shown centred on MM1/MM2 and MM4. A zoomed inset of MM1/MM2 is also shown in (a). Overplotted magenta contours in all panels show the 1.05\,mm continuum emission at 5$\sigma$. These contours are also shown in the zoomed insets, but at 5, 8, 16, 64 and $600\sigma$ (where $\sigma = 0.25$\,mJy\,beam$^{-1}$).  In panel (d), white contours show the VLA NH$_{3}$(3,3) peak intensity map (in $10\sigma$ steps from $5-145\sigma$, where $\sigma$ = 1.1\,mJy\,beam$^{-1}$, non-primary beam corrected), black contours show the VLA NH$_{3}$(6,6) peak intensity map (at the $5\sigma$ level, where $\sigma=0.9$\,mJy\,beam$^{-1}$, non-primary beam corrected), and the orange cross indicates the peak position of the candidate VLA 25\,GHz CH$_{3}$OH 5(2,3)-5(1,4) maser emission (see \S\ref{sec:vla_methanol}). The NH$_{3}$(3,3) maser groups from \S\ref{sec:vla_ammonia} are numbered in panel (d) as in Table~\ref{tab:nh333_group_properties}.  Overplotted white crosses in (e) mark the positions of 44\,GHz Class~I CH$_{3}$OH masers from \citet{cyganowski09}. Each panel shows the ALMA synthesised beam in the bottom left (as well as the VLA synthesised beam in panel (d)), a $1''$ scale bar in the bottom right, and the molecule name, frequency and upper energy in the top right.}
\label{fig:peakmaps}
\end{figure*}

Figure~\ref{fig:peakmaps} presents peak intensity maps of a selection of lines that are representative of the extended emission in our ALMA tuning.
These include three of the  lines targeted in our narrow spectral windows (N$_{2}$H$^{+}$(3--2), DCN v=0 ($4-3$) and H$_{2}$CO($4_{0,4}-3_{0,3}$), with E$_{u}/k_{B}=26.8$\,K, 34.8\,K and 34.9\,K respectively), and three from our wide spectral windows (H$_{2}$CO($4_{2,3}-3_{2,1}$), CH$_{3}$OH($6_{1,5}-5_{1,4}$) and CH$_{3}$OH($9_{-1,9}-8_{0,8}$), with E$_{u}/k_{B}=82.1$\,K, 63.7\,K and 110.0\,K respectively).  
The H$_{2}$CO($4_{2,3}-3_{2,1}$) and CH$_{3}$OH($6_{1,5}-5_{1,4}$) lines in Figure~\ref{fig:peakmaps}(c) and \ref{fig:peakmaps}(d) were identified as having extended emission around MM1 in \paperone{}. 

The four H$_{2}$CO and CH$_{3}$OH lines shown in Figure~\ref{fig:peakmaps} appear to spatially trace the same bi-polar outflow structure identified by \cite{cyganowski11a} in $^{12}$CO(2--1) with the SMA (see Figure~\ref{fig:g19}) and in HCO$^{+}$(1--0) and SiO(2--1) with CARMA. Kinematically, however, the ALMA H$_{2}$CO and CH$_{3}$OH emission traces lower velocity gas than the SMA $^{12}$CO or the CARMA HCO$^+$ emission, with a median full velocity extent of 25\,km\,s$^{-1}$ compared to $\sim$\,135\,km\,s$^{-1}$ for $^{12}$CO(2--1) and  $\sim$\,76\,km\,s$^{-1}$ for HCO$^{+}$(1--0) \citep{cyganowski11a}. 
The H$_2$CO and CH$_3$OH kinematics seen with ALMA do however exhibit a similar asymmetry as observed with the SMA and CARMA, with the blue-shifted lobe extending to higher velocities (up to $\sim$20\,km\,s$^{-1}$ from the systemic velocity, V$_{\mathrm{sys}}$) than the red-shifted lobe (up to $\sim$9\,km\,s$^{-1}$ from V$_{\mathrm{sys}}$). 
Overplotted in Figure~\ref{fig:peakmaps}(e) are the positions of 44\,GHz Class~I CH$_{3}$OH masers from \cite{cyganowski09}, which appear to trace the outer edge of the CH$_{3}$OH and H$_{2}$CO outflow lobes. 
Taken together, our results are consistent with the ALMA CH$_{3}$OH and H$_{2}$CO emission tracing lower velocity, outflow--cloud interaction regions or outflow cavity walls, as also seen in the EGO G11.92--0.61 by  \cite{cyganowski17}.

\begin{figure}
\centering
\includegraphics[scale=0.71]{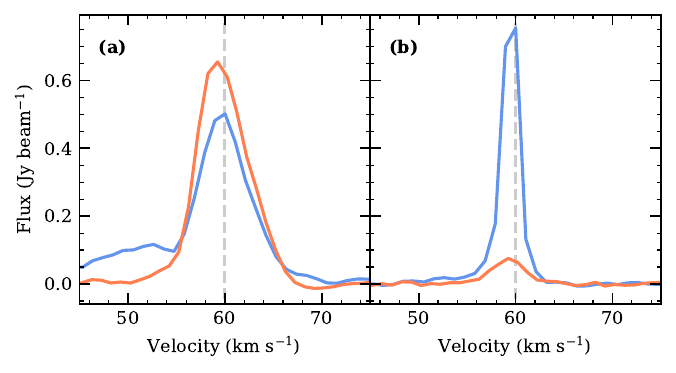}
\caption{Spectra of the CH$_{3}$OH$(9_{-1,9}-8_{0,8})$ line at $E_{u}/k_{B}=110.0$\,K (in blue) and the CH$_{3}$OH$(6_{1,5}-5_{1,4})$ line at at $E_{u}/k_{B}=63.7$\,K (in orange) towards (a) the ALMA 1.05\,mm continuum peak of MM1 (Table~\ref{tab:leaves}), and (b) the position of peak CH$_{3}$OH$(9_{-1,9}-8_{0,8})$ emission in the northern outflow lobe (\S\ref{sec:extended_emission}). The systemic velocity of MM1 \citep[$59.9\pm1.1$\,km\,s$^{-1}$;][]{cyganowski11a} is marked by the vertical dashed line.}
\label{fig:ch3oh_spectra}
\end{figure}

The CH$_{3}$OH($9_{-1,9}-8_{0,8}$) line in Figure~\ref{fig:peakmaps}(e) (with $\nu_{\mathrm{rest}}=278.30451$\,GHz) is known to exhibit Class~I maser emission towards other sources in the literature \citep[e.g.][]{voronkov12,yanagida14,cyganowski17}. 
Comparing the emission of this line with that of CH$_{3}$OH($6_{1,5}-5_{1,4}$) ($\nu_{\mathrm{rest}}=292.67291$\,GHz), shown in Figure~\ref{fig:peakmaps}(d), the two lines have similar emission morphologies. 
As shown in Figure~\ref{fig:ch3oh_spectra}, these lines also have similar fluxes towards the ALMA 1.05\,mm continuum peak of MM1 (equivalent to brightness temperatures of T$_{b}=39$ and 33\,K respectively).  
However, towards the region of brightest CH$_{3}$OH($9_{-1,9}-8_{0,8}$) emission in the outflow (T$_b=49$\,K at 18$^{\rm h}$25$^{\rm m}$44\fs517 $-$12$^{\circ}$22\arcmin32\farcs652 (J2000)), the CH$_{3}$OH($6_{1,5}-5_{1,4}$) line is an order of magnitude weaker (Figure~\ref{fig:ch3oh_spectra}) with T$_{b}=4$\,K. 
The CH$_{3}$OH($9_{-1,9}-8_{0,8}$) line is also notably narrower at the outflow position than towards the MM1 continuum peak ($\Delta$V$_{\rm FWHM}=2.01\pm0.01$\,km\,s$^{-1}$ and $6.95\pm0.08$\,km\,s$^{-1}$, respectively), and is spatially and kinematically coincident with 44\,GHz Class~I CH$_{3}$OH masers \citep{cyganowski09}, candidate 229.759\,GHz CH$_{3}$OH  maser emission \citep[][see their Fig.12]{cyganowski11a}, and NH$_{3}$(3,3) masers (see \S\ref{sec:vla_ammonia}).
The coincidence with probable 229.759\,GHz maser emission is notable because the 229.759 and 278.305\,GHz transitions are in the
same Class~I maser series \citep[the 36\,GHz maser series;][]{voronkov12}, and \cite{cyganowski18} found that all 229\,GHz masers in the EGO G11.92--0.61 have probable 278\,GHz maser counterparts.
Though the angular resolution  of our ALMA observations is insufficient to rely on line brightness temperatures to distinguish between thermal and maser behaviour \citep[as was also the case for G11.92--0.61 in][]{cyganowski17}, the line properties shown in Figure~\ref{fig:ch3oh_spectra}, and the coincidence of the brightest CH$_{3}$OH($9_{-1,9}-8_{0,8}$)  emission in the outflow with other shock-excited masers, strongly suggest thermalised emission towards the MM1 hot core, and 278\,GHz Class~I CH$_{3}$OH maser emission tracing outflow--cloud interaction regions \citep[e.g.][]{voronkov12,yanagida14,cyganowski17}.  Higher angular resolution observations of the CH$_{3}$OH($9_{-1,9}-8_{0,8}$) line would be required to confirm the presence of maser emission based on brightness temperature.

As a dense gas tracer, N$_{2}$H$^{+}$ is known to trace infrared-dark clumps at both quiescent and active evolutionary stages: \cite{sanhueza12}, for example, detect N$_{2}$H$^{+}$(1--0) towards the majority of their 92 infrared-dark clumps with the Mopra telescope at 38$''$ angular resolution (typically $\sim$0.8\,pc across their sample). Notably, while we observe extended N$_{2}$H$^{+}$(3--2) emission in G19.01--0.03 (Figure~\ref{fig:peakmaps}a), its morphology does not resemble that of the 1.05\,mm dust emission or that of the outflow structure.  
\cite{cyganowski17} found similar results in their ALMA N$_{2}$H$^{+}$(3--2) observations of the EGO G11.92--0.61 (their Fig.~3c), suggesting that in high-mass clumps the N$_{2}$H$^{+}$(3--2) and millimetre continuum emission do not trace the same structures on small size scales.\footnote{N$_{2}$H$^{+}$(3--2) is detected in absorption against the millimetre continuum of MM1, see Appendix~\ref{appendix:n2hp}.} We also note that larger-scale N$_{2}$H$^{+}$(3--2) emission will be affected by spatial filtering due to missing short-spacing information (Table~\ref{tab:obs}; \S\ref{sec:alma_obs}).
The morphology of the DCN(4--3) emission, on the other hand (Figure~\ref{fig:peakmaps}(f)), more closely resembles the ALMA 1.05\,mm continuum.   DCN(4--3) is detected towards MM1 to 38$\sigma$ (where $\sigma=6.3$\,mJy\,beam$^{-1}$):  the position of the DCN emission peak is offset from the ALMA 1.05\,mm continuum peak by 0.1$\arcsec$\,$\sim$\,400\,{\sc au} (within a DCN beam; Table~\ref{tab:obs_narrowline}), likely due to line opacity and self-absorption towards the continuum peak. The DCN emission also traces some of the nearby millimetre companions to MM1, which is discussed in Section~\ref{sec:mm2_mm5}.

\subsubsection{Ammonia emission with the VLA}
\label{sec:vla_ammonia}

Of the ammonia lines in our VLA tuning, NH$_{3}$(3,3) and NH$_{3}$(6,6) are known to exhibit maser behaviour thought to arise due to outflow-induced shocks \citep[e.g.][]{mangum94,kraemer95,brogan11}. Towards G19.01$-$0.03, spatially compact regions of NH$_{3}$(3,3) emission are detected at the $\geq$5$\sigma$ level (shown as white contours in Figure~\ref{fig:peakmaps}(d)) coincident both spatially and kinematically with the outflow-tracing ALMA H$_{2}$CO and CH$_{3}$OH emission shown in Figure~\ref{fig:peakmaps}(b)-(e). To characterise the NH$_{3}$(3,3) emission, we identify as firm detections, and potential maser candidates, locations where $\geq$5$\sigma$ emission in $\geq$2 consecutive velocity channels is spatially contiguous within a VLA beam.  Following the terminology of \cite{towner21}, we refer to emission in a single velocity channel (each 0.4\,km\,s$^{-1}$ wide; Table~\ref{tab:obs_narrowline}) as a ``spot'', and emission from multiple spatially contiguous spots as a ``group'': a group thus contains at least two spots.

\begin{table*}
\centering
\caption{Properties of NH$_{3}$(3,3) emission groups.}
\label{tab:nh333_group_properties}
\setlength\tabcolsep{9.9pt}
\begin{tabular}{ccccccccc}
\hline\hline
Group$^{a}$ & \multicolumn{4}{c}{J2000.0 centroid position$^{b}$}                                                     & I$_{\mathrm{peak}}$$^{c}$  & Angular spread $^{d}$          & V$_{\mathrm{min}}$, V$_{\mathrm{max}}$\,$^{e}$          & V$_{\mathrm{peak}}$\,$^{f}$                \\ \cline{2-5}
ID    & $\alpha$ ($^{\mathrm{h\,m\,s}}$) & $dx$ ($''$) & $\delta$ ($^{\circ}$ $'$ $''$) & $dy$ ($''$)        & (mJy beam$^{-1}$)        & $\alpha$, $\delta$ ($\arcsec$, $\arcsec$)    & (km\,s$^{-1}$)       & (km\,s$^{-1}$)       \\ \hline
1     & 18:25:44.89        & 0.02                & -12:23:01.55        & 0.02                 & 20.1 (1.2)        &    0.012, 0.044                       &  60.2, 61.4                    & 60.6                     \\
2     & 18:25:44.79        & 0.02                & -12:23:01.02        & 0.02                 & 18.9 (1.3)        &    0.073, 0.091                       &  59.8, 61.4                    & 60.2                     \\
3     & 18:25:44.85        & 0.03                & -12:22:37.44        & 0.03                 & 9.5  (1.2)        &    0.016, 0.022                       &  57.4, 57.8                    & 57.4                     \\
4      & 18:25:44.49        & 0.01                & -12:22:32.60        & 0.01                 & 216.0 (1.3)       &    0.262, 0.059                       &  51.4, 61.8                    & 59.8                     \\
5     & 18:25:44.53        & 0.01                & -12:22:30.54        & 0.01                 & 59.0  (1.3)       &    0.092, 0.030                       &  57.8, 59.4                    & 59.0                     \\
6     & 18:25:45.11        & 0.02                & -12:22:29.89        & 0.02                 & 26.2  (1.3)       &    0.048, 0.019                       &  59.0 , 59.4                    & 59.4                     \\
7    &  18:25:45.69       & 0.03                &  -12:22:29.73       & 0.03                 & 12.4  (1.3)       &    0.039, 0.067              &  58.6, 59.4                    & 59.4                     \\ 
8    & 18:25:44.60        & 0.03                & -12:22:27.85        & 0.03                 & 11.9  (1.3)       &    0.041, 0.037                       &  59.4, 59.8                    & 59.8                     \\ \hline
\end{tabular}
\begin{flushleft}
    $^a$  Labelled from south to north, see Figure~\ref{fig:peakmaps}(d).\\
    $^b$  Intensity-weighted centroid position, and intensity-weighted mean statistical error from the Gaussian fitting, of each emission group.\\
    $^c$  Fitted peak intensity of the brightest emission spot in each group (i.e. from a single channel). The statistical error from the Gaussian fitting in listed in parentheses. $T_{B}$(K) $\approx 7186 \times I$(Jy~beam$^{-1}$).\\
    $^d$  The angular spread of the emission group, calculated as the standard deviation of the difference between the position of each individual emission spot (i.e. from a single channel) and the intensity- weighted centroid position of the group. \\
    $^e$  Velocity spread over which $\geq5\sigma$ emission is present, see \S\ref{sec:vla_ammonia}.\\
    $^f$  Velocity of the channel at which the brightest emission in each group appears.
\end{flushleft}
\end{table*}

\begin{figure}
\centering
\includegraphics[scale=0.5]{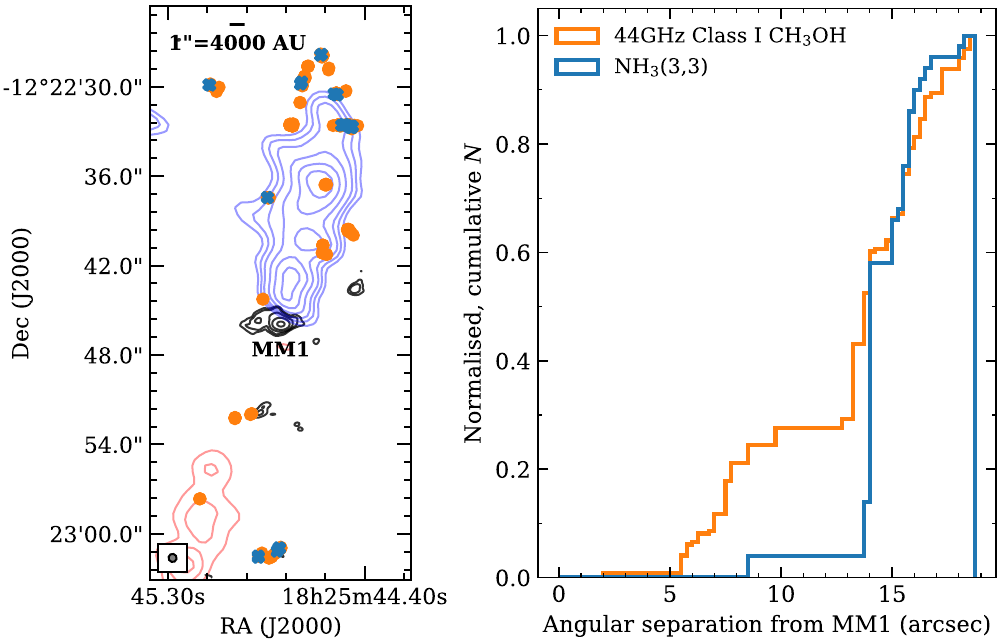}
\caption{\textit{Left panel:} Fitted positions of each NH$_{3}$(3,3) spot across all groups (blue $\times$), and 44\,GHz Class~I CH$_{3}$OH maser spot positions from \citet{cyganowski09} (orange filled circles). ALMA 1.05\,mm continuum contours are plotted in black (at [5, 8, 16, 64, 600]$\times\sigma$, where $\sigma = 0.25$\,mJy\,beam$^{-1}$). High-velocity, blue- and red-shifted $^{12}$CO(2--1) emission observed with the SMA \citep[][]{cyganowski11a} are plotted in blue and red contours, as in Figure~\ref{fig:g19}. The VLA beam and a scalebar are plotted in the bottom left and top left of the panel respectively. \textit{Right panel:} Normalised, cumulative histogram of the angular separation (in arcseconds) of each NH$_{3}$(3,3) spot (blue), and each 44\,GHz Class~I CH$_{3}$OH maser spot (orange), from the ALMA 1.05\,mm continuum peak of MM1 (Table~\ref{tab:leaves}). Angular separations were calculated using the \texttt{au.angularSeparation} task of the analysisUtils {\sc python} package.}
\label{fig:maser_separations}
\end{figure}

To extract the position, peak intensity and velocity of each $\geq5\sigma$ emission spot, 
we fit the observed emission with a 2D Gaussian using the {\sc casa imfit} task \citep[as in][]{towner21}.
The rms noise is measured in each channel to ascertain the signal-to-noise of the detection 
as the noise is higher in channels with bright emission
due to dynamic range limitations. 
As we expect the emission to be unresolved, we fit each emission spot as a point source by fixing the size 
to that of the synthesised beam \citep[e.g.][]{hunter18, towner21}. 
In total, we identify 50 emission spots at the $\geq5\sigma$ level (where the mean $\sigma$ across all channels with emission is $\sigma=1.14$\,mJy\,beam$^{-1}$) that reside in 8 emission groups. The position of each emission spot is plotted in Figure~\ref{fig:maser_separations}, each maser group is labelled in Figure~\ref{fig:peakmaps}(d), and Table~\ref{tab:nh333_group_properties} lists the properties of each group.
Unlabelled contours in Figure~\ref{fig:peakmaps}(d) correspond to emission that does not meet our criteria of $\geq$5$\sigma$ emission in $\geq$2 consecutive velocity channels within a VLA beam (generally because the emission centroids shift by more than a beam in consecutive channels) and so is not included in our analysis.
The strongest emission group (ID 4 in Table~\ref{tab:nh333_group_properties}) has a fitted peak intensity ($216.0\pm1.3$\,mJy\,beam$^{-1}$) corresponding to a brightness temperature $\sim$1550\,K, strongly suggestive of masing behaviour. 
Four additional emission groups  (i.e. IDs 1, 2, 5 and 6) also exhibit brightness temperatures greater than the $E_{u}/k_{B}$ of the line.
While the brightest group (ID 4) spans both blue- and red-shifted velocities (Table~\ref{tab:nh333_group_properties}), blue-shifted and red-shifted NH$_{3}$(3,3) emission groups generally reside towards the blue- and red-shifted outflow lobes respectively.  The same pattern was observed in 44\,GHz Class~I CH$_{3}$OH  masers by \citet{cyganowski09} (their Figure~5f), though the maser velocities are modest compared to the velocity extent of the thermal molecular outflow lobes \citep{cyganowski09,cyganowski11a}. 
All NH$_{3}$(3,3) emission spots are coincident both spatially and kinematically (within a beam and a channel, respectively) with 44\,GHz Class~I CH$_{3}$OH maser spots from \citet{cyganowski09} (see Figure~\ref{fig:maser_separations}), similar to the results seen in the EGO G35.03$+$0.35 by \cite{brogan11}.
We note that in G19.01$-$0.03, not all 44\,GHz methanol masers are coincident with NH$_{3}$(3,3) masers.

In G19.01$-$0.03, all identified NH$_{3}$(3,3) emission spots are located in the outer portions of the bipolar outflow lobes, whilst the 44\,GHz Class~I methanol maser spots are also found closer to the driving source.
This distinction is shown in Figure~\ref{fig:maser_separations}, which plots the normalised, cumulative distribution of the angular separation ($\theta_\mathrm{sep}$) of each NH$_{3}$(3,3) and 44\,GHz Class~I CH$_{3}$OH maser spot from the ALMA 1.05\,mm dust continuum peak of MM1.
The two-sample Kolmogorov-Smirnov (K-S) and $k$-sample Anderson-Darling (A-D) tests are both nonparametric null hypothesis tests often used to compare two such distributions. The null hypothesis in this case states that the two distributions may be drawn from the same underlying distribution if the tests return $p$-values larger than 0.05. Running both the K-S and A-D tests on the full range of angular separations ($\theta_\mathrm{sep}$) shown in Figure~\ref{fig:maser_separations} returns vanishingly small $p$-values, meaning the null hypothesis may be rejected and a conclusion drawn that the NH$_{3}$(3,3) and 44\,GHz methanol maser distributions are significantly different. However, running both tests for $\theta_\mathrm{sep}\geq12.5\arcsec$ returns $p\gg0.05$, meaning that the two distributions at large angular separations are effectively indistinguishable and may be drawn from the same underlying distribution.

We note that thermal NH$_{3}$(1,1) and (2,2) emission are detected towards G19.01--0.03 at 73$\arcsec$-resolution by \citet{cyganowski13} but are undetected in our VLA observations (to $5\sigma$ limits of 6.2 and 5.9\,mJy\,beam$^{-1}$ respectively).
Our non-detection of these low-J transitions is primarily attributable to the limited brightness temperature sensitivity of our high-resolution observations (Table~\ref{tab:obs_narrowline}), with spatial filtering potentially contributing to the non-detection of the NH$_{3}$(2,2) line (the maximum recoverable scale of our observations is only $\sim$4.5$\arcsec$\,$\sim$0.08\,pc at D=4\,kpc, see Table~\ref{tab:obs}). 
\citet{cyganowski13} also detect NH$_{3}$(3,3) towards G19.01--0.03 that is likely dominated by thermal emission, presumably arising near MM1. Extended NH$_{3}$(3,3) emission is also seen towards the mm continuum sources in the EGO G35.03$+$0.35 by \cite{brogan11} in $3.7\arcsec\times3.0\arcsec$ resolution VLA observations \citep[$\sim$~$8,580\times6,960$\,{\sc au} at the parallax distance of 2.32$^{+0.24}_{-0.20}$\,kpc from][]{wu14}.  
In our data, NH$_{3}$(3,3) emission at $\ge$5$\times$ the median rms in Table~\ref{tab:obs_narrowline} is present in a single channel within the area of MM1's mm continuum emission. This channel, however, has bright maser emission and so an elevated rms, and the NH$_{3}$(3,3) peak is $<$5$\times$ the channel rms.  Since the NH$_{3}$(3,3) peak is also offset from the mm continuum peak and our tentative NH$_{3}$(6,6) and (7,7) detections (see below), we conclude that NH$_{3}$(3,3) is undetected towards MM1 in our data.
NH$_{3}$(5,5) is also undetected in our VLA observations, to a $5\sigma$ limit of 5.5\,mJy\,beam$^{-1}$. 

NH$_{3}$(6,6) emission (shown in black contours in Figure~\ref{fig:peakmaps}(d)) is detected at the $5.7\sigma$ level (I$_{\rm peak}=$5.4\,mJy\,beam$^{-1}$) towards a position in the outskirts of the northern, blue-shifted outflow lobe (18$^{\rm h}$25$^{\rm m}$44\fs50 $-$12$^{\circ}$22\arcmin32\farcs52 (J2000)), spatially and kinematically coincident with an NH$_{3}$(3,3) maser group (ID 4 in Table~\ref{tab:nh333_group_properties}), 44\,GHz Class~I CH$_{3}$OH masers \citep{cyganowski09}, and candidate ALMA 278.3\,GHz Class~I CH$_{3}$OH maser emission (Section~\ref{sec:extended_emission}). 
Towards MM1, NH$_{3}$(6,6) is detected at the $5.6\sigma$ level (I$_{\rm peak}=$5.3\,mJy\,beam$^{-1}$), spatially and kinematically coincident with a tentative $4.4\sigma$ (4.6\,mJy\,beam$^{-1}$) detection  of NH$_{3}$(7,7). These tentative high-J detections towards MM1 could represent tentative evidence of emission from hot gas originating in the inner regions of the circumstellar accretion disc around the central MYSO(s) (\paperone{}).

\subsubsection{25\,GHz methanol emission with the VLA}
\label{sec:vla_methanol}

\begin{figure*}
\centering
\includegraphics[scale=0.57]{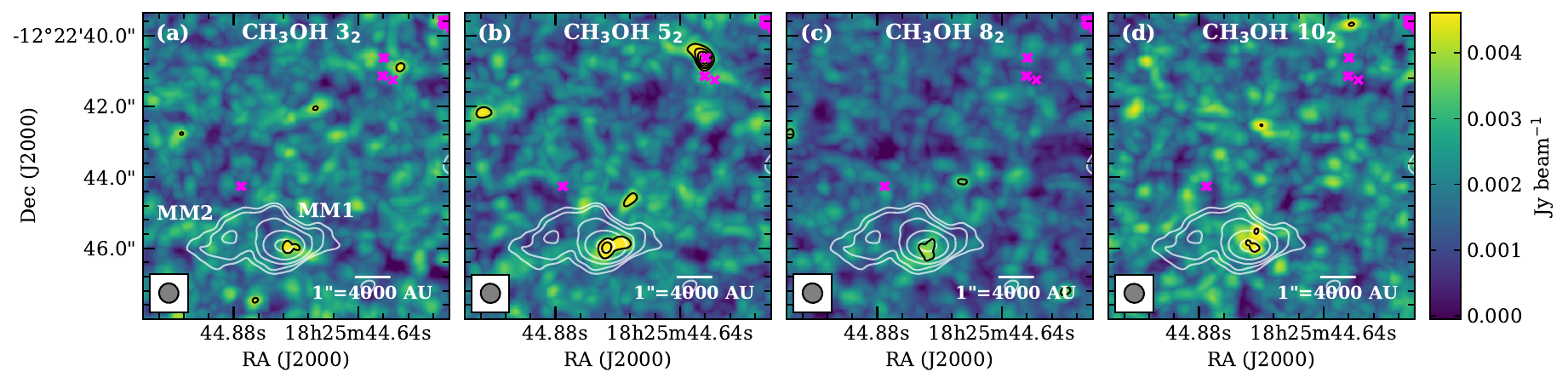}
\caption{VLA peak intensity maps of the four $\sim$25\,GHz CH$_{3}$OH lines in our VLA tuning (Table~\ref{tab:obs_narrowline}), here referred to by the short hand notation 3$_2$, 5$_2$, 8$_2$ and 10$_2$. Each panel shows the peak intensity map of the labeled transition in colourscale and black contours, with contour levels of 4$\sigma$ for the 3$_2$, 8$_2$ and 10$_2$ lines (where $\sigma = 1.04$, 0.85 and 1.19\,mJy\,beam$^{-1}$ respectively), and [4, 5, 6, 7]$\times\sigma$ for the 5$_{2}$ line (where $\sigma = 1.01$\,mJy\,beam$^{-1}$).  ALMA 1.05\,mm continuum contours are overplotted in white at 5, 8, 16, 64 and 600$\sigma$ (where $\sigma=0.25$\,mJy\,beam$^{-1}$).  Magenta crosses mark the positions of 44\,GHz Class~I CH$_{3}$OH masers from \citet{cyganowski09}. Each synthesised beam, and a $1\arcsec$ scale bar, are plotted in the bottom left and bottom right of each panel respectively.}
\label{fig:25ghz_mom8}
\end{figure*}

The four CH$_{3}$OH lines in our VLA tuning (Table~\ref{tab:obs_narrowline}) can exhibit thermal and/or Class~I maser emission and are commonly detected towards EGOs:
in their study of 20 EGOs,
\cite{towner17} detected thermal and/or maser emission towards 16 (80\%) of their sources.
As \cite{towner17} adopt a 4$\sigma$ detection criterion, we consider $\ge$4$\sigma$ 25 GHz CH$_3$OH emission so that we can compare to their results; we also adopt their shorthand notation of 3$_2$, 5$_2$, 8$_2$ and 10$_2$ for these lines.  
Towards MM1, we tentatively detect all four lines at the 4.8, 5.5, 4.5 and 4.5$\sigma$ levels, respectively (see Figure~\ref{fig:25ghz_mom8}, and Table~\ref{tab:obs_narrowline} for  $\sigma$ values), equivalent to brightness temperatures ($T_{B}$) of 32.6, 36.8, 23.2 and 32.4\,K.  \cite{towner17} detect all four 25 GHz CH$_3$OH lines in half of their sample (i.e. $\sim$63\% of those with a detection of at least one line).
The strongest emission towards MM1 is in the 5$_{2}$ line, a result also reported by \cite{towner17} towards the majority of the sources in their sample. 
We also detect 7.8$\sigma$ 5$_{2}$ emission ($T_{B}=52.3$\,K) offset by $6\arcsec\sim24,000$\,{\sc au} to the north-west of MM1 (see Figures~\ref{fig:peakmaps}(d) and \ref{fig:25ghz_mom8}).  This 5$_{2}$ emission is positionally and kinematically coincident with outflow-tracing H$_{2}$CO and CH$_{3}$OH emission observed with ALMA (\S\ref{sec:extended_emission} and Figure~\ref{fig:peakmaps}) and with 44\,GHz Class~I CH$_{3}$OH maser emission \citep{cyganowski09}.

\begin{figure*}
\centering
\includegraphics[scale=0.46]{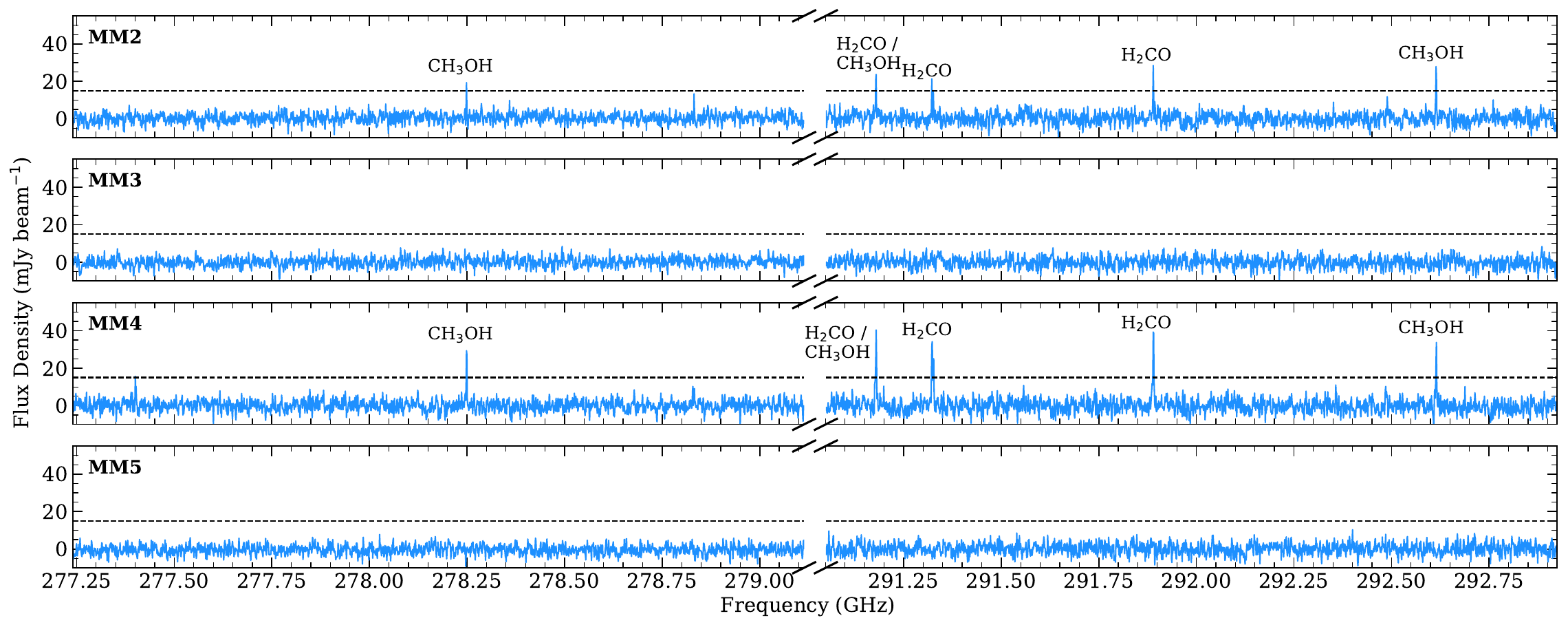}
\caption{ALMA spectra towards the 1.05\,mm continuum peaks of MM2, MM3, MM4 and MM5 (Table~\ref{tab:leaves}) for the two broad spectral windows. The break in the frequency axis between the two broad spectral windows is marked by the bold black lines. The $5\sigma$ level (where $\sigma=3$\,mJy\,beam$^{-1}$, \S\ref{sec:alma_obs}) is marked by the horizontal dashed black line. Lines with emission $>5\sigma$ are labelled.}
\label{fig:neighbours}
\end{figure*}

To distinguish between thermal and maser 25 GHz CH$_3$OH emission, we consider the two approaches detailed by \cite{towner17}: (i) following criteria on the spatial and spectral extent of the line emission, and (ii) comparing observed line intensities to predictions for optically thin LTE emission.  For the first approach, spatially and spectrally broad emission (i.e. resolved $\geq4\sigma$ emission in $\geq5$\,channels $\sim2$\,km\,s$^{-1}$) is considered thermal, whilst unresolved/point-like spectrally narrow emission (i.e. $\geq4\sigma$ emission in $\leq4$\,channels $\sim1.6$\,km\,s$^{-1}$) is considered a candidate maser. The strongest 5$_{2}$ emission towards the MM1 outflow appears both unresolved (i.e. with spatial extent less than a beam) and spectrally narrow with a fitted full-width half-maximum velocity width ($\Delta V_{\rm FWHM}$) of $0.6\pm0.1$\,km\,s$^{-1}$. We therefore consider this a candidate maser.
Towards MM1, the emission from all four 25\,GHz CH$_3$OH lines appears spatially unresolved. 
Both broad and narrow Gaussian components are needed to describe the 5$_{2}$ emission towards MM1, with velocity widths of $3.5\pm0.1$ and $0.7\pm0.1$\,km\,s$^{-1}$ respectively, while the spectral profiles of the 3$_{2}$, 8$_{2}$ and 10$_{2}$ lines are best fit by single Gaussians with velocity widths of $3.4\pm0.1$, $6.0\pm0.2$ and $6.7\pm0.2$\,km\,s$^{-1}$ respectively. 
All of the fitted Gaussian components 
have centroid velocities consistent with the systemic velocity of MM1 \citep[59.9$\pm$1.1\,km\,s$^{-1}$;][]{cyganowski11a}.  The 8$_{2}$ and 10$_{2}$ lines also have velocity widths of a similar order to the thermal CH$_{3}$OH emission seen with ALMA towards MM1 (see left panel of Figure~\ref{fig:ch3oh_spectra}).
Line models under LTE and optically thin conditions (produced using the Weeds package of the {\sc class} software) reasonably describe the broad emission components of all four lines towards MM1.  As a whole, our data suggest that both thermal (broad components) and weak maser (narrow 5$_{2}$ component) may be present towards MM1, though we emphasise that this is a tentative result due to the low S/N of the 25\,GHz CH$_3$OH emission towards MM1.

\subsubsection{MM2...MM5 with ALMA and the VLA}
\label{sec:mm2_mm5}

In Figure~\ref{fig:neighbours}, we show spectra for the broad ALMA spectral windows towards the 1.05\,mm continuum peaks of MM2, MM3, MM4 and MM5. The lines shown in Figure~\ref{fig:peakmaps}(c)-(e) lie within these broad bands. 
MM3 and MM5 appear devoid of molecular line emission in these bands, whilst MM2 and MM4 both have $>5\sigma$ detections of a handful of lines (labelled in Figure~\ref{fig:neighbours}). All of the line emission detected towards MM2 and MM4 in these broad bands appears spatially and kinematically coincident with emission attributed to the outflow driven by MM1 (including the H$_{2}$CO and CH$_{3}$OH lines shown in Figures~\ref{fig:peakmaps}(c), (d) and (e); see \S\ref{sec:extended_emission}). 
As such, it is likely that this line emission is associated with the MM1 outflow, rather than being physically associated with MM2 and MM4.  
N$_{2}$H$^{+}$(3--2) emission (see Figure~\ref{fig:peakmaps}(a)) is coincident with all four new millimetre sources, however given the emission does not morphologically resemble the ALMA 1.05\,mm continuum, it is unlikely to arise from the same volumes as MM2...MM5.
As shown in Figure~\ref{fig:peakmaps}(f), DCN($4-3$) does exhibit an emission morphology similar to the ALMA 1.05\,mm continuum. MM2 and MM4 are detected in emission at the 10$\sigma$ and 9$\sigma$ levels respectively (where $\sigma=6.3$\,mJy\,beam$^{-1}$), and do not exhibit signs of self-absorption (unlike MM1, see \S\ref{sec:extended_emission}), whilst MM3 and MM5 are undetected to a $5\sigma$ limit of 31.5\,mJy\,beam$^{-1}$. The positions of peak DCN emission towards MM2 and MM4 are offset from their corresponding ALMA 1.05\,mm continuum peaks by 0.09 and 0.40$\arcsec$ respectively (equivalent to 360 and 1600\,{\sc au}), within a DCN beam ($\sim$0.47$\arcsec$; see Table~\ref{tab:obs_narrowline}).
Systemic velocities for MM2 and MM4 are found from Gaussian fitting of the DCN peak emission to be $59.88\pm0.06$\,km\,s$^{-1}$ and $59.72\pm0.05$\,km\,s$^{-1}$ respectively, consistent with the $59.9\pm1.1$\,km\,s$^{-1}$ systemic velocity of MM1 \citep{cyganowski11a}, kinematically linking MM2 and MM4 to MM1 and the G19.01$-$0.03 clump.

A handful of Class~I 44\,GHz CH$_{3}$OH maser spots lie $0.60\arcsec\sim2400$\,{\sc au} to the east of the ALMA 1.05\,mm continuum peak of MM4 (see Figure~\ref{fig:peakmaps}(e)), coincident spatially and kinematically with H$_{2}$CO and CH$_{3}$OH emission attributed to the MM1 outflow (see \S\ref{sec:extended_emission}). 
No NH$_{3}$($J$\,=\,$K$\,=\,1, 2, 3, 5, 6, 7) emission is detected towards, or in the immediate vicinity of, MM2...MM5 in our VLA data, to $5\sigma$ limits of 6.2, 5.9, 5.7, 5.5, 4.8, 5.3\,mJy\,beam$^{-1}$ respectively. Similarly, MM2...MM5 are undetected in the 25\,GHz CH$_{3}$OH lines in our VLA observations ($5\sigma$ limits 5.2, 5.1, 4.3 and 6.0\,mJy\,beam$^{-1}$ for the 3$_2$, 5$_2$, 8$_2$ and 10$_2$ lines).  The CH$_{3}$OH 5$_{2}$ emission spot detected towards the outflow driven by MM1 (see \S\ref{sec:vla_methanol}) is $3.6\arcsec$\,$\sim$\,$14,200$\,{\sc au} away from the ALMA 1.05\,mm continuum peak of MM3.
In all, with no clear evidence for protostellar activity in MM2...MM5 in our ALMA and VLA data, the current evidence suggests that these sources may be starless condensations, prestellar cores, millimetre knots in MM1's outflow lobes, or a combination of the three (see also Section~\ref{sec:low-mass}).

\section{Discussion}
\label{sec:discussion}

\subsection{Physical properties of MM1 from \texorpdfstring{CH$_{3}$OH emission}{}}
\label{sec:all_mm1_ch3oh}

In the following section, we estimate the column density and temperature of MM1 from the observed CH$_{3}$OH emission using two complementary methods: (i) a rotation diagram analysis (\S\ref{sec:rotdiag}), and (ii) a set of synthetic spectra that represent the observed emission (\S\ref{sec:synthspec}).

\subsubsection{Rotation diagram analysis of CH$_{3}$OH}
\label{sec:rotdiag}

Assuming local thermodynamic equilibrium and optically thin emission, a single temperature may describe the emission of all observed lines. This temperature is defined by the Boltzmann distribution, written as:
\begin{equation}
    \log \left(\frac{N_{u}}{g_{u}}\right) = \log \left(\frac{N}{Q(T_{\mathrm{rot}})}\right) - \frac{E_{u}}{k_{B}T_\mathrm{{rot}}} \,,
    \label{eq:boltz_lte}
\end{equation}
\noindent where $N_{u}$ is the upper energy level column density, $g_{u}$ is the upper energy state degeneracy, $N$ is the total column density, $Q(T_{\mathrm{rot}})$ is the partition function, $E_{u}$ is the upper level energy, $k_{B}$ is the Boltzmann constant, and $T_{\mathrm{rot}}$ is the rotational temperature \citep[e.g.][]{goldsmithlanger99}. The left-hand side of equation~\ref{eq:boltz_lte} may be re-written  as:
\begin{equation}
    \log\left(\frac{N_{u}}{g_{u}}\right) = \log\left(\frac{3k_{B}\int T_{\mathrm{mb}} \mathrm{dV}}{8\pi^{3}\nu S_{_{ij}}\mu^{2}}\right) \,,
    \label{eq:boltz_lte_lhs}
\end{equation}
\noindent where $\int T_{\mathrm{mb}}\mathrm{dV}$ is the integrated line strength over velocity with units of $\mathrm{K}\times\mathrm{km\,s^{-1}}$, $\nu$ is the rest frequency of the transition, $\mu$ is the permanent dipole moment with units of Debye, and $S_{_{ij}}$ is the unit-less intrinsic line strength \citep[e.g.][]{cummins86,herbst09}. 
When multiple lines of a species are detected, assuming that their emission arises from the same physical structure, and assuming that all lines are optically thin, one may construct a rotation diagram by plotting $\log(N_{u}/g_{u})$ against $E_{u}/k_{B}$. Fitting a straight line will yield $T_{\mathrm{rot}}$ from the slope, and $N/Q(T_{\mathrm{rot}})$ from the y-intercept. If a straight line reasonably fits the rotation diagram, the assumption is made that all transitions are thermalised, and thus $T_{\mathrm{rot}}$ is expected to equal the excitation temperature, $T_{\mathrm{ex}}$ \citep[e.g.][]{goldsmithlanger99}. Taking this $T_{\mathrm{rot}}$, the total column density ($N$) may be found with the partition function, which following \cite{townes_schawlow55} and \cite{purcell09}, for example, may be written for CH$_{3}$OH as: 
\begin{equation}
    Q(T_{\mathrm{rot}})=1.2327\,T_{\mathrm{rot}}^{1.5} \,\, .
\end{equation}

\begin{figure}
\centering
\includegraphics[scale=.65]{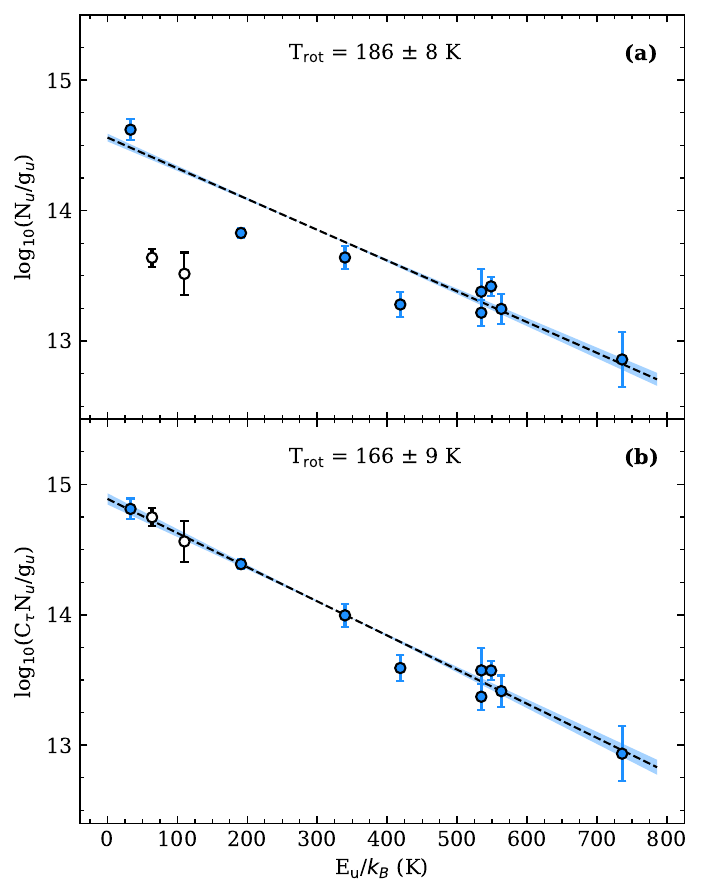}
\caption{Rotation diagram of CH$_{3}$OH at the ALMA 1.05\,mm dust continuum peak of MM1, (a) before opacity correction, and (b) following opacity correction. The best fit is shown by the dashed black line, with the statistical uncertainty on the slope and y-intercept represented by the shaded blue region. The CH$_{3}$OH($6_{1,5}-5_{1,4}$) and CH$_{3}$OH($9_{-1,9}-8_{0,8}$) lines (with  $E_{u}/k_{B}=63.7$ and $110.0$\,K respectively) are coloured in white as they are outflow tracers (see Figure~\ref{fig:peakmaps}) and optically thick.}
\label{fig:rotdiag}
\end{figure}

The most detected species towards MM1 in our spectral tuning at $>$5$\sigma$ is CH$_{3}$OH, with thirteen transitions identified across both the wide and narrow bands (see \S\ref{sec:lines}) with $E_{u}/k_{B}$ ranging $32-736$\,K. 
We produce a rotation diagram towards the ALMA 1.05\,mm continuum peak of MM1 (Table~\ref{tab:leaves}) using eleven CH$_{3}$OH transitions, excluding the CH$_{3}$OH($15_{1,0}-14_{2,0}$) line from the wide bands, and CH$_{3}$OH($6_{1,2} - 5_{1,2}$) from the narrow bands, as they are significantly blended by the H$_{2}$CO(4$_{2,3}$--3$_{2,2}$) and CH$_{3}$OCHO($27_{1,27} - 26_{0,26}$) lines respectively.
We include both A and E transitions in this analysis since we do not detect enough transitions to to fit them separately. The integrated intensity of each line is evaluated from Gaussian profiles fitted to the CH$_{3}$OH spectra at the ALMA 1.05\.mm continuum peak position (see Table~\ref{tab:leaves}). 
Errors on the Gaussian fits are propagated through to the rotation diagram, and the best fit is found following a weighted least-squares fit. 
This rotation diagram is shown in Figure~\ref{fig:rotdiag}(a), where it can be seen that two data points (marked white, belonging to the CH$_{3}$OH($6_{1,5}-5_{1,4}$) and CH$_{3}$OH($9_{-1,9}-8_{0,8}$) lines with E$_{u}/k_{B}=63.7$ and 110.0\,K respectively) deviate from the general trend of the other nine lines. These lines are seen in Figure~\ref{fig:peakmaps} to be outflow tracing, and their generally low $\log\left(N_{u}/g_{u}\right)$ is an indication that they may also be optically thick. 
The derived rotational temperature is $186\pm8$\,K.

As previously stated, a rotation diagram is constructed under the assumption that all lines are optically thin. Optically thick lines can inflate derived rotational temperatures due to an artificially shallow slope \citep[e.g.][]{goldsmithlanger99}. We counter this by applying an opacity correction, $C_{\tau_{i}} = \tau_{i}/(1 - e^{\tau_{i}})$, where the subscript $i$ refers to the $i^{\mathrm{th}}$ CH$_{3}$OH transition. We follow \cite{brogan07,brogan09} by iteratively solving for the $\tau_{i}$ and $T_{\mathrm{rot}}$ values that produce the best fit, minimising the $\chi^{2}$ value. Figure~\ref{fig:rotdiag}(b) shows the best fitting opacity-corrected rotation diagram. The $\tau_{i}$ values for this best fit range between $\sim0.4-11.7$, with indeed the CH$_{3}$OH($6_{1,5}-5_{1,4}$) and CH$_{3}$OH($9_{-1,9}-8_{0,8}$) lines being the most optically thick (with $\tau=11.7$ and 11.3 respectively). The highest-J, CH$_{3}$OH($23_{4,19}-22_{5,18}$) line with E$_{u}/k_{B}=736$\,K is unsurprisingly the most optically thin.  
Of the lines with compact emission morphologies (i.e. that trace MM1 and not the outflow) in the 278~GHz wideband spectral window, the CH$_{3}$OH($14_{4,10}-15_{3,12}$) line is the most optically thick, with $\tau=1.9$, implying an emitting region of $\sim$0\farcs25$\sim$ 1000\,AU \citep[following Section 4.4 of][]{brogan09}.  
The derived opacity-corrected rotational temperature is $166\pm9$\,K, and the column density is $(2.0\pm0.4)\times10^{18}$\,cm$^{-2}$.

\subsubsection{Synthetic CH$_{3}$OH spectra}
\label{sec:synthspec}

We make a second estimate of the column density and temperature of MM1 by producing synthetic spectra that are representative of the observed data using the \textsc{weeds} extension \citep{maret11} of the \textsc{class} software. \textsc{weeds} solves the radiative transfer equation assuming a state of local thermodynamic equilibrium (LTE), and takes into account the background continuum emission.  Using values for the frequency, Einstein A coefficients, partition function, and upper level degeneracy and energy from the JPL and CDMS catalogues \citep{jpl, cdms}, Weeds calculates the opacity of each line, meaning that corrections such as line broadening are applied to optically thick lines. 
The free parameters of the synthetic spectra are the column density, excitation temperature, systemic velocity, FWHM velocity width, and source size. The projected diameter of the telescope is also required to calculate the beam dilution/filling factor.  We refer the reader to \cite{maret11} for a more detailed description of the full procedure.

We explore the posterior distribution of the free parameters of the LTE line models using the Python package \textsc{emcee} \citep{emcee}, an affine-invariant Monte Carlo Markov Chain (MCMC) sampler \citep{goodman10}. A posterior distribution of model parameters consistent with the data is evaluated by multiple ``walkers'', allowed to walk for a number of iterations. A ``burn-in'' period of $n_{\mathrm{burn}}$ iterations allows the walkers to converge on the posterior region. These ``burn-in'' solution chains are then discarded before beginning a ``production'' run with walkers initialised around the posterior region identified during the burn-in. Walkers are allowed to walk through the posterior for $n_{\mathrm{prod}}>n_{\mathrm{burn}}$ iterations, to allow probing of the covariance of the model parameters. The model parameters that are most representative of the data are found following the maximisation of the log-likelihood function:
\begin{equation}
    \mathcal{L}(T_{\mathrm{mb}}|\theta) = -0.5 \sum_{i}\left( \frac{T_{\mathrm{mb},i}^{\mathrm{mod}} - T_{\mathrm{mb},i}^{\mathrm{obs}}}{\sigma^{\mathrm{obs}}_{\mathrm{rms}}} \right)^{2} \, ,
\end{equation}
\noindent where $\theta$ are the model parameters, $T_{\mathrm{mb},i}^{\mathrm{mod}}$ and $T_{\mathrm{mb},i}^{\mathrm{obs}}$ are the model and observed main-beam brightness temperature of the $i^{\mathrm{th}}$ channel, respectively, and $\sigma^{\mathrm{obs}}_{\mathrm{rms}}$ is the observed rms noise of the spectrum (see \S\ref{sec:alma_obs}).  The statistical uncertainty on each model parameter is evaluated from the 0.16 and 0.84 quantiles of the posterior (i.e. the 1$\sigma$ level). Our scripts are publicly available\footnote{\href{2}{https://github.com/gwen-williams/WeedsPy\_MCMC}}, and are packaged together in a reusable form we call \textsc{weedspy\_mcmc} where they are generalised to take in user-defined parameters such as the priors, number of walkers, initial walker positions within the parameter space, and the number of burn-in and production iterations, from a text file. 
A {\sc class}-readable text file is created containing the $\theta$ model parameters currently walked to and the name of the molecule to be modelled, and then a {\sc class} script is called that runs the LTE model for those $\theta$. Plots are generated of the highest likelihood synthetic spectrum, the corner plot of the 1D and 2D posterior distributions of the free parameters (as in Figure~\ref{fig:corner}(a)), and the trace plot of the \textsc{emcee} walkers through the parameter space. 
To efficiently sample the probability distributions for the free parameters of the synthetic spectra generated with \textsc{weeds}, spectral windows are treated separately in our scripts.

\subsubsection{Pixel-by-pixel analysis of CH$_{3}$OH synthetic spectra}
\label{sec:synthspec_results}

Applying this procedure to our ALMA data, we find the highest likelihood model on a pixel-by-pixel basis across MM1, allowing investigation of any spatial variation in the column density, temperature, centroid velocity and velocity width. As in \S\ref{sec:rotdiag}, this analysis focuses on CH$_{3}$OH emission, as it is the species with the highest number of identified transitions in our tuning. Of the thirteen total CH$_{3}$OH lines in the tuning, seven are located in the $\sim$278\,GHz band, four are in the $\sim$292\,GHz band (see Figure~\ref{fig:spec6} and Table~\ref{tab:lines}), one lies in the narrow spectral window that targets N$_{2}$H$^{+}$($3-2$), and one lies in the narrow spectral window that targets DCN($4-3$). 
As the CH$_{3}$OH($6_{1,5}-5_{1,4}$) and CH$_{3}$OH($9_{-1,9}-8_{0,8}$) lines (with $\nu_{\mathrm{rest}}=292.67291$ and 278.30451\,GHz respectively) are shown in Figure~\ref{fig:peakmaps} to be outflow tracing, we exclude them from this analysis since an assumption made here is that the emitting region is the same for all modelled transitions. We also exclude the CH$_{3}$OH($15_{1,0}-14_{2,0}$) line ($\nu_{\mathrm{rest}}=291.24057$\,GHz) as it is heavily blended by a close-by H$_{2}$CO line (see also \S\ref{sec:rotdiag}), the CH$_{3}$OH($6_{1,2}-5_{1,2}$) line ($\nu_{\mathrm{rest}}=289.62430$\,GHz) as it is blended with a CH$_{3}$OCHO line (see \S\ref{sec:lines} and \S\ref{sec:rotdiag}), and the CH$_{3}$OH($10_{1,10}-9_{0,9}$) line ($\nu_{\mathrm{rest}}=292.51744$\,GHz) as it is the only identified line in the first torsionally excited state (i.e. $v_t = 1$). 
This leaves 6 lines in the $\sim$278\,GHz band, 1 line in the $\sim$292\,GHz band, and 1 line in a narrow band. 
Since modelling one line is not sufficient to constrain four free parameters, we conduct our MCMC analysis on the six remaining CH$_{3}$OH lines in the $\sim$278\,GHz band (indicated in bold in Table~\ref{tab:lines}).  We then use the best-fit parameters to generate synthetic spectra for the other spectral windows and check the resulting spectra for goodness of fit (see also Section~\ref{sec:abundances_estimation}).

\begin{figure*}%
\includegraphics[width=.69\textwidth]{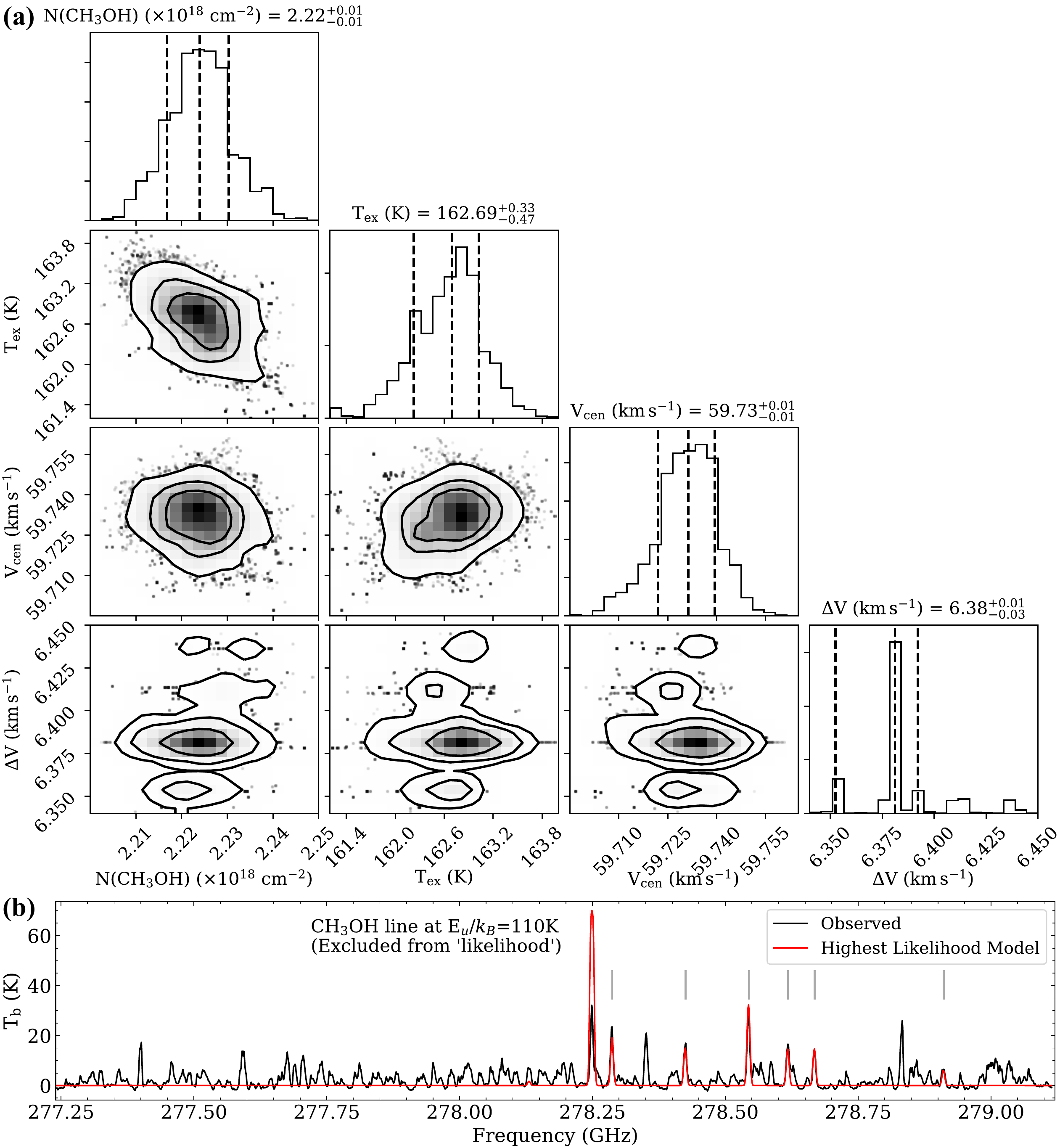}%
\hfill
\begin{minipage}[b]{.29\textwidth}%
\caption{(a) One-dimensional and two-dimensional histograms of the posterior distributions for the free parameters in the LTE synthetic spectral modelling (N(CH$_{3}$OH), T$_{\mathrm{ex}}$, $\mathrm{V}_{\mathrm{cen}}$ and $\Delta \mathrm{V}$) towards the ALMA 1.05\,mm continuum peak. Black contours on the 2D histograms are placed at 2, 1.5 and 1\,$\sigma$, whilst the dashed black lines on the 1D histograms are placed at the 0.16, 0.5 and 0.84 quantiles. (b) Observed spectrum towards the ALMA 1.05\,mm continuum peak (black) overplotted with the corresponding highest likelihood model (red) i.e. model parameters defined by the mode of the posterior, printed in panel (a). The six CH$_{3}$OH lines that are modelled and included in the maximisation of the log-likelihood function are marked by vertical grey lines. The excluded CH$_{3}$OH($9_{-1,9}-8_{0,8}$) line with $E_{u}/k_{B}=110.0$\,K is labelled.}
\label{fig:corner}
\end{minipage}%
\end{figure*}

\begin{figure*}
\centering
\includegraphics[scale=.5425]{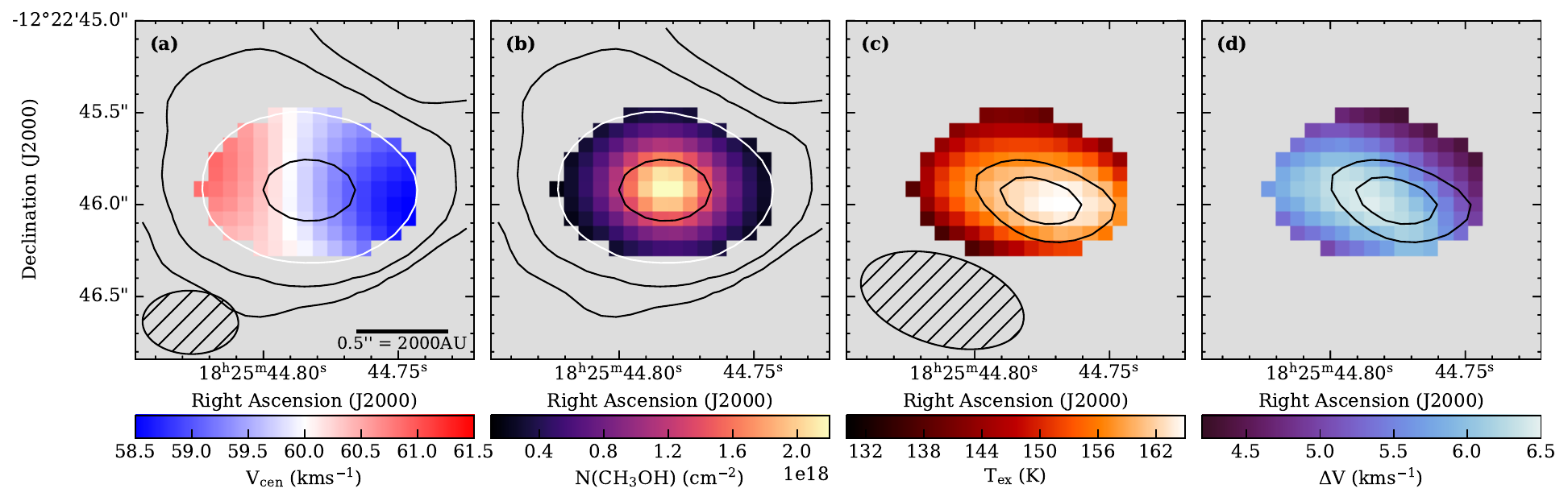}
\caption{Pixel-by-pixel highest likelihood model parameters from the Weeds CH$_{3}$OH synthetic spectra towards MM1: (a) centroid velocity in km\,s$^{-1}$, (b) CH$_{3}$OH column density in cm$^{-2}$, (c) excitation temperature in Kelvin, and (d) FWHM velocity width in km\,s$^{-1}$. ALMA 1.05\,mm continuum contours are plotted in panels (a) and (b) at 8, 16, 64 and 200$\sigma$ (where $\sigma = 0.25$\,mJy\,beam$^{-1}$). The white contour (at $64\sigma$) represents the level above which pixels were fed into \textsc{weedspy\_mcmc}, and matches the source size of MM1 (see Table~\ref{tab:leaves}). Contours of VLA 5.01\,cm emission (from \paperone{}) are plotted in panels (c) and (d) at the 4 and 5$\sigma$ levels (where $\sigma=5.0$\,$\mu$Jy\,beam$^{-1}$) in panel. A scale bar and the ALMA beam are plotted in panel (a), and the VLA beam is plotted in panel (c).}
\label{fig:best_lte}
\end{figure*}

For our pixel-by-pixel modelling, we fix the source size to the geometric mean of the ${\sim}$278\,GHz synthesised beam (${\sim}0.5\arcsec$).\footnote{The image cube is converted to brightness temperature using the Rayleigh-Jeans approximation and beamsize with the \texttt{tt.brightnessImage} function of \texttt{toddTools}.}
We also limit the pixel area being modelled to within the continuum source size of MM1 from Table~\ref{tab:leaves} (see white contour in Figures~\ref{fig:best_lte}(a) and (b)), shown in \paperone{} to match well the region of compact molecular line emission  (see their Fig.3).  
The projected diameter of the ALMA array was calculated to be 558\,metres\footnote{Calculated using the \texttt{au.getBaselineLengths} function from the analysisUtils Python package}, and we estimate the background continuum on a pixel-by-pixel basis \citep{mcguire18}, with the ALMA 1.05\,mm continuum converted to brightness temperature\footnote{Conversion done under the Rayleigh-Jeans approximation using the \texttt{tt.brightnessImage} function of \texttt{toddTools}.}. We initialise 60 walkers, and allow them to run for $n_{\mathrm{burn}}=200$ and $n_{\mathrm{prod}}=500$ iterations (see Appendix~\ref{appendix:emcee} for a discussion of these choices). 
The remaining free parameters are the column density, excitation temperature, systemic velocity and FWHM velocity width. We assume uniform, uninformative priors, and initialise the walkers in a four-dimensional sphere about an initial value that lies within those priors. Our priors are estimated from initial ``dummy'' runs of the LTE synthetic spectra outside of \textsc{weedspy\_mcmc}, and are set to be  $(0.05-3.0)\times10^{18}$\,cm$^{-2}$ for the column density, $80-300$\,K for the temperature, $56.0 - 64.0$\,km\,s$^{-1}$ for the centroid velocity, and $2.0 - 8.0$\,km\,s$^{-1}$ for the velocity width.

Figure~\ref{fig:corner}(a) shows the 1D and 2D posterior distributions of column density, temperature, centroid velocity and velocity width, evaluated for the spectrum at the ALMA 1.05\,mm continuum peak of MM1. The narrow, normally distributed  1D posteriors for column density, temperature and centroid velocity show that these parameters are reasonably constrained.
The slope in the 2D column density--temperature posterior distribution indicates an unsurprising, but only slight, degeneracy between these two parameters, meaning the statistical errors are small despite this. There is no obvious covariance of the centroid velocity with either column density or temperature. The velocity width is less well constrained, with two slight peaks around the mode of the posterior distribution (within  $<0.03$\,km\,s$^{-1}$). 
Though significantly smaller than the spectral resolution of our data ($\sim$1\,km\,s$^{-1}$), this uncertainty is attributed to line blending rather than under-sampling as the double peak persists in tests where $n_{\mathrm{prod}}$ is increased to $>$3000. The double peak also persists when walkers are initialised with a different random number seed, so is independent of initial walker position. The highest likelihood model spectrum at the continuum peak position of MM1 is shown in Figure~\ref{fig:corner}(b), with parameters N$($CH$_{3}$OH$)=(2.22\pm0.01)\times10^{18}$\,cm$^{-2}$, T$_{\mathrm{ex}}=162.7\substack{+0.3 \\ -0.5}$\,K, V$_{\mathrm{cen}}=59.73\pm0.01$\,km\,s$^{-1}$ and $\Delta V = 6.38\substack{+0.01 \\ -0.03}$\,km\,s$^{-1}$.  These remain unchanged (within the errors) when walkers are initiated with a different random number seed, and when $n_{\mathrm{prod}}$ is increased to $>3000$. These column density and temperature values are remarkably consistent with those derived from the opacity-corrected rotation diagram of $(2.0\pm0.4)\times10^{18}$\,cm$^{-2}$ and $166\pm9$\,K respectively (see \S\ref{sec:rotdiag}), and the centroid velocity is also consistent with the known systemic velocity of $59.9\pm1.1$\,km\,s$^{-1}$ \citep{cyganowski11a}. 
The line optical depths returned by {\sc class} are lower than those estimated in the rotation diagram analysis with an average $\tau=0.3$ compared to $\tau=1.0$ (excluding CH$_{3}$OH($9_{-1,9}-8_{0,8}$) with $E_{u}/k_{B}=110.0$\,K).
We attribute the higher line opacities estimated from the rotation diagram to the source size not being fixed in the rotation diagram approach (see \S\ref{sec:rotdiag}).
Figure~\ref{fig:corner}(b) shows the synthetic spectrum well reproduces the observed CH$_{3}$OH emission, other than significantly over-estimating the line strength of the optically thick, outflow tracing CH$_{3}$OH($9_{-1,9}-8_{0,8}$) line (see Figure~\ref{fig:peakmaps}) that was excluded from the maximisation of the log-likelihood. This vindicates its exclusion from the analysis, further suggesting that the emission from the CH$_{3}$OH($9_{-1,9}-8_{0,8}$) line may not occur in the same volume as the other modelled, optically thin lines.

\subsubsection{Images of CH$_{3}$OH physical parameters}
\label{sec:synthspec_images}

Maps of the pixel-by-pixel, highest likelihood CH$_{3}$OH model parameters towards MM1 are shown in Figure~\ref{fig:best_lte}.
The highest likelihood centroid velocity map (Figure~\ref{fig:best_lte}a) shows a clear gradient in velocity across MM1 in agreement with the gradient reported in \paperone{} from a simpler moment analysis of individual lines. 
The CH$_{3}$OH column density (Figure~\ref{fig:best_lte}b) peaks coincident with, and follows the same morphology as, the ALMA 1.05\,mm continuum emission.
The position of peak CH$_{3}$OH temperature (18$^{\rm h}$25$^{\rm m}$44\fs763 $-$12$^{\circ}$22\arcmin46\farcs00 (J2000); see Figure~\ref{fig:best_lte}c) is however offset from the ALMA 1.05\,mm dust continuum and CH$_{3}$OH column density peaks by $0.22\arcsec\sim880$\,{\sc au}. 
Whilst the column density and temperature are somewhat degenerate (see Figure~\ref{fig:corner}a), the difference between the offset peak temperature ($165.5\pm0.6$\,K) and the temperature at the column density peak ($162.7\pm0.4$\,K) is significantly greater than its propagated statistical error (i.e. $2.8\pm0.7$\,K). In \paperone{}, a centimetre counterpart to MM1 (called CM1) was detected for the first time with the VLA at 5.01 and 1.21\,cm. The VLA 1.21\,cm emission peak was coincident with the ALMA 1.05\,mm continuum peak of MM1, whilst the VLA 5.01\,cm emission peak of CM1 (18$^{\rm h}$25$^{\rm m}$44\fs773 $-$12$^{\circ}$22\arcmin46\farcs00 (J2000); \paperone{}) was found to be offset from MM1 by 0.16$\arcsec\sim640$\,{\sc au} (\paperone{}). 
Indeed, the entire morphology of the region of elevated CH$_{3}$OH temperature matches that of the VLA 5.01\,cm emission, with both being elongated along a south-easterly direction relative to MM1 (as shown in Figure~\ref{fig:best_lte}c). 
The velocity width map in Figure~\ref{fig:best_lte}(d) also shows a similar elongation to the VLA 5.01\,cm continuum, however it does not peak coincident with either the CH$_{3}$OH column density, the CH$_{3}$OH temperature, nor the VLA 5.01\,cm continuum position.
Following analysis of the combined centimetre-millimetre spectral energy distribution of CM1/MM1 in \paperone{}, it was revealed that a free-free emission component was required to describe the VLA 5.01\,cm emission. This component was interpreted as a small, 66\,{\sc au} diameter gravitationally trapped hypercompact (HC) H{\sc ii} region, which directly implies the presence of a source causing ionisation and heating of its surroundings.  Given that the morphology of the elevated CH$_{3}$OH temperature aligns with that of the VLA 5.01\,cm continuum, and that the position of peak CH$_{3}$OH temperature is only offset from the VLA 5.01\,cm position of CM1 by $0.07\arcsec\sim280$\,{\sc au} (equivalent to the absolute positional uncertainty of the VLA 5.01\,cm data of 0.07$\arcsec$; \paperone{}), it is reasonable to surmise that both are attributable to the same underlying source/mechanism. We further speculated in \paperone{} that the VLA 5.01\,cm emission being misaligned with respect to the direction of the bipolar outflow emanating from MM1 was suggestive of the ionisation being driven by a second object, perhaps indicative of an unresolved high-mass binary system. In that case, a second possible origin for the offset elevated CH$_{3}$OH temperature could be the presence of an ionised jet driven by the unresolved high-mass binary companion. Outflows and jet-like outflows are noted in the literature to cause heating of molecular gas \citep[e.g.][]{zhang07,wang12_outflow}.

\subsection{Physical properties of MM1...MM5 from dust emission}
\label{sec:low-mass}

As mentioned in \S\ref{sec:intro} and \S\ref{sec:synthspec_images}, it was shown in \paperone{} from an analysis of the centimetre-millimetre spectral energy distribution of MM1 that its ALMA 1.05\,mm flux was 99.99 per cent dominated by thermal dust emission.  MM2...MM5 are also likely dominated by thermal dust emission, given their non-detection in the centimetre at 1.21 and 5.01\,cm with the VLA to 5$\sigma$ limits of 30 and 25$\mu$Jy\,beam$^{-1}$ respectively (\paperone{}). 
As such, the masses ($\mathrm{M_{\mathrm{gas}}}$) of MM1...MM5 may be calculated from their integrated flux densities at 1.05\,mm ($S_{1.05\,\mathrm{mm}}$) assuming isothermal dust emission following \cite{hildebrand83}:
\begin{equation}
	\mathrm{M_{\mathrm{gas}}} = \frac{d^{2}\,R\,S_{1.05\,\mathrm{mm}}\,C_{\tau_{\mathrm{dust}}}}{\kappa_{1.05\,\mathrm{mm}}\,B_{\nu}(T_{\mathrm{dust}})} \, ,
	\label{eq:mdust}
\end{equation}
\noindent where $d$ is the distance, $R$ is the gas-to-dust mass ratio (here assumed to be 100), $\kappa_{1.05\,\mathrm{mm}}$ is the dust opacity at $1.05\,\mathrm{mm}$, $B_{\nu}(T_{\mathrm{dust}})$ is the Planck function, and $T_{\mathrm{dust}}$ is the dust temperature. We include a correction for dust optical depth, $C_{\tau_{\mathrm{dust}}}$ as:
\begin{equation}
    C_{\tau_{\mathrm{dust}}}= \tau_{\mathrm{dust}}/(1-e^{-\tau_{\mathrm{dust}}}) \, ,
\end{equation}
\noindent where $\tau_{\mathrm{dust}}$ may be estimated as:
\begin{equation}
    \tau_{\mathrm{dust}} = -\ln\left(1-\frac{T_{b}}{T_{\mathrm{dust}}}\right) \, .
\end{equation}
\noindent We estimate $T_{b}$ as the mean brightness temperature across the size of each source (as listed in Table~\ref{tab:leaves}), though this may underestimate $\tau_{\mathrm{dust}}$ for small-scale structures. 
From \cite{ossenkopf94}, we set $\kappa_{1.05\,\mathrm{mm}}=1.45$\,cm$^{2}$g$^{-1}$ for dust grains with thick ice mantles in high density gas \citep[as in \paperone{};][]{cyganowski17}. 
We also assume that the dust is well-coupled to the gas such that $T_{\mathrm{dust}} = T_{\mathrm{gas}}$. The temperatures we assume for each of the sources is discussed below in \S\ref{sec:mm1_mass} and \S\ref{sec:other_masses}. Assuming spherical geometry, a mean molecular weight per hydrogen molecule ($\mu_{\mathrm{H}_{2}}$) of 2.8 \citep[e.g.][]{kauffmann08}, and a radius equal to half the geometric mean of the source sizes presented in Table~\ref{tab:leaves}, we further calculate the average H$_{2}$ column density and H$_{2}$ volume density. All derived source properties are listed in Table~\ref{tab:cont_masses}.

\subsubsection{Temperature, mass and stability of MM1}
\label{sec:mm1_mass}

In \paperone{}, we assumed a temperature range for MM1 of $100-130$\,K based on the CH$_{3}$CN(J$=$12--11) line fitting of \cite{cyganowski11a} from 2.3$\arcsec$ angular resolution SMA data. Our new ALMA data (with 0.4$\arcsec$ angular resolution)  improves on that by a factor of $\sim$30 in beam area. We revise our temperature assumption to 163\,K, as calculated at the ALMA 1.05\,mm continuum peak of MM1 in \S\ref{sec:synthspec_results} from the CH$_{3}$OH synthetic spectra. This is an increase of 33--66\,K on the SMA-derived temperature. Observations of the EGO G11.92--0.61 (hereafter G11.92) by both \cite{cyganowski11a} and \cite{ilee16} provide us with a point of comparison on temperature. With $2.4\arcsec$ angular resolution SMA observations, \cite{cyganowski11a} derived 77 and 166\,K temperatures for G11.92 from a two-component fit to CH$_{3}$CN emission. Taking the same approach but with $0.5\arcsec$ angular resolution SMA observations (a factor of $\sim$27 improvement in beam area), \cite{ilee16} find the two G11.92 temperature components to be 150 and 230\,K. This is an increase of 60--70\,K in temperature, attributed to the difference in probed spatial scales between the two data sets. The same resolution-dependent behaviour is observed here for the temperature of G19.01 MM1.

\begin{table}
\centering
\caption{Derived properties for mm continuum sources.}
\label{tab:cont_masses}
\setlength\tabcolsep{8.0pt}
\begin{tabular}{ccccccc}
\hline\hline
Source & T$_{b}$ & T$_{\mathrm{dust}}$ & $\tau_{\mathrm{dust}}$ & M$_{\mathrm{gas}}$     & N(H$_{2}$)        & n(H$_{2}$)        \\
       &         &                     &                        &                        & $\times10^{23}$   & $\times10^{6}$    \\
       & (K)     & (K)                 &                        & (M$_{\odot}$)          & (cm$^{-2}$)       & (cm$^{-3}$)       \\ \hline
MM1    & 10.9    & 163               & 0.07                   & 4.3                     & 6.7               & 17.0              \\
MM2    & 3.5     & 24                & 0.16                   & 1.3                     & 3.2               & 10.3              \\
       &         & 15                & 0.27                   & 2.7                     & 6.6               & 21.1              \\
MM3    & 3.0     & 24                & 0.13                   & 1.0                     & 1.5               & 3.9               \\
       &         & 15                & 0.22                   & 2.1                     & 3.1               & 7.8               \\
MM4    & 2.9     & 24                & 0.13                   & 0.9                     & 1.4               & 3.4               \\ 
       &         & 15                & 0.22                   & 1.8                     & 2.7               & 6.8               \\
MM5    & 2.8     & 24                & 0.13                   & 0.1                     & --                & --                \\
       &         & 15                & 0.21                   & 0.2                     & --                & --                \\
\hline
\end{tabular}
\begin{justify}
Column 1: Source name. Col. 2: Planck brightness temperature averaged over the source size listed in Table~\ref{tab:leaves}. Col. 3: Assumed dust temperature, derived from the LTE synthetic spectra for MM1 (see \S\ref{sec:synthspec_results}), and a temperature range of $15-24$\,K assumed for MM2...MM5 based on large scale clump observations \citep{schuller09,cyganowski13,elia17}. Col. 4: Dust optical depth, calculated from the average $T_{b}$ in column 2. Col. 5: Opacity-corrected gas mass, in solar masses. Col. 6: H$_{2}$ column density, and Col. 7: H$_{2}$ volume density, both assuming spherical symmetry.
\end{justify}
\end{table}

The mass of MM1 is here revised, from M$_{\mathrm{gas}}=5.4-7.2$\,M$_{\odot}$ at 130--100\,K derived in \paperone{}, to M$_{\mathrm{gas}}=4.3$\,M$_{\odot}$ at 163\,K (Table~\ref{tab:cont_masses}). Attributing this gas mass to the now known circumstellar disc around MM1, the stability of the disc can be assessed by estimation of the disc-to-star mass ratio (as done in \paperone{}), with unstable discs exhibiting typical values of $M_{\mathrm{gas}}/M_{\ast}>0.1$ \citep[e.g.][]{kratter16}. Following \paperone{}, the stellar mass ($M_{\ast}$) can be calculated as $M_{\ast}=M_{\mathrm{enc}}-M_{\mathrm{gas}}$ where $M_{\mathrm{enc}}$ is the enclosed mass within the disc outer radius (estimated as $\sim40-70$\,M$_{\odot}$ from kinematic modelling in \paperone{}). The disc-to-star mass ratio was found in \paperone{} to be $\sim0.08-0.22$ hinting that the disc could be unstable and be undergoing fragmentation into as-of-yet undetected low-mass stellar companions. It was noted however that should the temperature of MM1 increase on the size-scales probed by the ALMA data \citep[as it did for the SMA observations of G11.92;][]{cyganowski11a,ilee16}, then the resulting lower $M_{\mathrm{gas}}$ could push the disc towards stability. Our revised $M_{\mathrm{gas}}/M_{\ast}$ is 0.07--0.12, indeed indicative of the disc being potentially stable against fragmentation. This approach does however risk underestimating the disc mass, as it does not account for variations in the temperature, dust opacity and dust optical depth in the disc \citep{johnston15,forgan16}.

\subsubsection{Low mass protocluster members}
\label{sec:other_masses}

As there are insufficient molecular lines detected with ALMA towards MM2...MM5 for the estimation of temperature, we use clump-scale  observations of G19.01 to inform our temperature assumptions. \cite{elia17} derive $T_{\mathrm{dust}}=17.9$\,K from SED fitting of HI-GAL observations, whilst \cite{schuller09} and \cite{wienen12} derive $T_{\mathrm{gas}}=19.5$\,K from NH$_{3}$ hyperfine structure fitting of $\sim$40$\arcsec$ angular resolution Effelsberg 100\,m data.  \cite{cyganowski13} also estimated temperature for the G19.01 clump using NH$_{3}$ observations, taken with the Nobeyama 45\,m telescope. Their single-temperature fit to the NH$_{3}$ spectra yielded a kinetic temperature of $23.8\pm0.4$\,K, but they found that a two-component fit with T$_{\rm kin, cool}$\,=\,$14.7\pm0.5$\,K and T$_{\rm kin,warm}$\,=\,$50.2\pm4.4$\,K better represented the observed data \citep{cyganowski13}.  As the warm $\sim$50\,K component is likely attributable to the central MYSO (MM1), we exclude it and use the remaining temperature estimates to inform an assumed temperature range for MM2...MM5 of $15-24$\,K. Table~\ref{tab:cont_masses} lists derived properties for $T_{\mathrm{dust}}=$\,15 and 24\,K.  We note that H$_{2}$ column and volume densities are not included for MM5 as its size is smaller than a beam.

The gas masses of MM2...MM5 range between $0.1-1.3$\,M$_{\odot}$ and $0.2-2.7$\,M$_{\odot}$ at 24 and 15\,K respectively (see Table~\ref{tab:cont_masses}), and their median radius is $\sim$2000\,{\sc au} (excluding the unresolved MM5; see Table~\ref{tab:leaves}). As such, the physical properties of these millimetre sources appear consistent with typical low-mass pre/protostellar cores observed in infrared dark clouds \citep[with masses and radii of a few solar masses and a few thousand {\sc au} respectively, e.g.][]{sanhueza19,morii21,redaelli22} rather than discs around low-mass YSOs \citep[with masses and radii $<$0.1\,M$_{\odot}$ and a few hundred {\sc au} respectively, e.g.][]{huang18,dullemond18}.
Furthermore, their lack of COM emission or signs of outflow activity indicate that they may be candidate low-mass pre-stellar rather than protostellar cores. 
Interestingly, this result contrasts with the situation in the EGO G11.92$-$0.61, where several of the low-mass cores identified by \citet{cyganowski17} drive molecular outflows, clearly identifying them as protostellar cores.

The number and spatial distribution of the low-mass cores in G19.01$-$0.03 also contrasts with G11.92$-$0.61, where \citet{cyganowski17}'s 1.05\,mm ALMA observations of the ATLASGAL clump detected 16 new compact mm continuum sources (for a total of 19, including 3 previously-known massive sources).  Compared to G11.92$-$0.61, the low-mass sources detected in G19.01$-$0.03 are located closer to the most massive member of the protocluster (median projected separation $\sim$0.08\,pc compared to $\sim$0.17\,pc) and comprise a lower fraction of the clump mass \citep[$<$1\% compared to 3-5\%;][]{cyganowski17}.
We consider our results in the context of expectations for thermal Jeans fragmentation. 
For the simple assumption of a uniform density sphere, the minimum mass necessary for a fragment to be bound, and so collapse, is
\begin{equation}
    M_J = \left( \frac{5\,c_s^2}{2 \,G}\right)^{3/2} \left(\frac{3}{4\pi\rho}\right)^{1/2} \, ,
\end{equation}
\noindent where $\rho$ is the density, and $c_{s}$ is the sound speed defined as:
\begin{equation}
    c_{s} = \left ( \frac{k_{B}\,T}{\mu_{H_{2}}\,m_{H}} \right )^{1/2}\, ,
\end{equation}
\noindent where $\mu_{H_{2}}$ is the mean molecular weight per hydrogen molecule (equal to 2.8; see also \S\ref{sec:low-mass}), and $m_{H}$ is the mass of a Hydrogen atom. 
The length scale for a fragment to be bound, under the same assumptions, is the Jeans radius
\begin{equation}
    R_J = \left(\frac{15\,c_s^2}{8\pi\,G \rho}\right)^{1/2}\, ,
\end{equation}
where $R_J$ is the radius of a sphere of mass
$M_J$ and the Jeans length ($\lambda_J$) is $\lambda_J\approx2R_{J}$. Using the Hi-GAL clump properties \citep[1165\,M$_{\odot}$ and R$=$0.175\,pc when scaled to D$=$4\,kpc, T$=$17.9\,K; \S\ref{sec:intro},][]{elia17}, we calculate the thermal Jeans radius and thermal Jeans mass of the G19.01$-$0.03 clump to be $R_{J}\approx0.012$\,pc and $M_{J}\approx0.36$\,M$_{\odot}$ respectively.  Using the ATLASGAL clump properties \citep[926\,M$_{\odot}$ and R$=$0.358\,pc for D$=$4.0kpc, T$=$19.5\,K, \S\ref{sec:intro},][]{schuller09}, $R_{J}\approx0.041$\,pc and $M_{J}\approx1.36$\,M$_{\odot}$.
The linear resolution of our ALMA data is $\sim$0.008\,pc, and the $5\sigma$ detection limit (converted to mass) is $0.1-0.3$\,M$_{\odot}$ at $24-15$\,K, so we are sensitive to the relevant fragmentation scales. 
With median masses of $1.0-2.0$\,M$_{\odot}$ at $24-15$\,K, respectively, MM2...MM5 are of order the thermal Jeans mass, consistent with the results of \citet{palau15} from their study of mm fragments within the inner 0.1\,pc of $\sim$20 massive dense cores.  The number of millimetre sources detected in G19.01 is however notably lower than expected for thermal Jeans fragmentation.  For the 13\% core formation efficiency (CFE) found by \citet{palau15}, \begin{math} N_{\rm Jeans}= ({M_{\rm clump} CFE)}/{M_{J}}\end{math} $\sim$420 and $\sim$90 for the Hi-GAL and ATLASGAL-based estimates, respectively.
Strong magnetic fields are expected to suppress fragmentation, and recent observational work found a tentative correlation between the number of mm fragments and the mass-to-flux ratio for a sample of 18 massive dense cores with polarization data \citep[][and references therein]{palau21}.  Observations of polarized dust emission in G19.01$-$0.03 are needed to assess whether a dynamically important magnetic field contributes to the low level of observed fragmentation.

\subsubsection{Virial analysis of MM4}
\label{sec:boundedness}

The virial parameter ($\alpha_{vir}$) is a commonly used diagnostic in determining the boundedness of structures. To qualify as a stellar progenitor, a source must be in a gravitationally bound state. This is satisfied by $\alpha_{vir}<2$ in the absence of pressure terms such as magnetic energy density. Furthermore, a core is said to be in hydrostatic equilibrium when $\alpha_{vir}\sim1$, or to be gravitationally unstable when $\alpha_{vir}<1$. The virial parameter is often expressed as follows:
\begin{equation}
    \alpha_{vir} = \frac{M_{vir}}{M_{\mathrm{core}}} \,,
\end{equation}
where $M_{vir}$ is the virial mass, and $M_{\mathrm{core}}$ is the observed core mass. The virial mass for a spherical core may be calculated following \cite{maclaren88}:
\begin{equation}
    M_{vir} = 3\left (\frac{5-2n}{3-n}\right) \frac{\sigma_{NT}^{2}R}{\mathrm{G}} \,,
    \label{eq:mvir}
\end{equation}
\noindent where $n$ is the index of the density profile of the source (where $\rho \propto  r^{-n}$), $R$ is the radius, and G is the gravitational constant. The non-thermal velocity dispersion, $\sigma_{NT}$, may be calculated following:
\begin{equation}
    \sigma_{NT}^{2} = \sigma_{\mathrm{obs}}^{2} - \sigma_{\mathrm{th}}^{2} = \sigma_{\mathrm{obs}}^{2} - \frac{k_{B}\,T_{\mathrm{dust}}}{\mu\,m_{H}} \,,
    \label{eq:sigmaNT}
\end{equation}
\noindent where $\sigma_{\mathrm{obs}}$ and $\sigma_{\mathrm{th}}$ are the observed and thermal velocity dispersions respectively \citep{bertoldimckee92}, 
$T_{\mathrm{dust}}$ is the dust temperature (assumed to equal the gas temperature), $\mu$ is the molecular weight of the molecular tracer, and $m_{H}$ is the mass of a Hydrogen atom.

As discussed in \S\ref{sec:extended_emission} and \S\ref{sec:mm2_mm5}, of the cold and dense gas tracers we detect it is only the emission from DCN (with $\mu$ = 28) that appears morphologically similar to the ALMA 1.05\,mm dust continuum emission, with $\geq5\sigma$ detections towards MM1, MM2 and MM4. As we cannot safely disentangle the DCN emission of MM2 from the surrounding envelope emission (see Figure~\ref{fig:peakmaps}), we focus our virial analysis on MM4 as it appears isolated.  
As we aim to test whether MM4 could be a prestellar core, we assume a flat density profile where $n=0$ \citep[e.g.][]{ward-thompson94}. DCN exhibits nuclear quadrupole hyperfine structure, with seven lines in the $J=4-3$ transition with velocity offsets from the main component between -1.6 to 2.1\,km\,s$^{-1}$ \citep{cdms} with a mean velocity offset of 0.1\,km\,s$^{-1}$ (within our 1\,km\,s$^{-1}$ spectral resolution). The effect of these components is most important in the lower energy DCN transitions such as $J=1-0$ or $J=2-1$ \citep[e.g.][]{parise09}, with the strongest central component dominating in transitions such as $J=4-3$. We fit the hyperfine structure lines on a pixel-by-pixel basis across MM4 using the \texttt{hfs} fitting routine of the {\sc class} software. 
We set $\sigma_{\mathrm{obs}}$ to the mean velocity dispersion across the FWHM size of MM4 of $0.58\pm0.03$\,km\,s$^{-1}$.
Taking the radius as half the geometric mean of the deconvolved size of MM4 listed in Table~\ref{tab:leaves} (equal to $\sim$2000\,AU), the virial mass of MM4 is calculated to be $3.8\pm0.4$\,M$_{\odot}$. 
With $M_{\mathrm{core}}$ calculated from the 1.05\,mm dust emission (1.84\,M$_{\odot}$ at 15\,K, or 0.91\,M$_{\odot}$ at 24\,K; see Table~\ref{tab:cont_masses}), this yields $\alpha_{vir}=2.0-4.1(\pm 0.4)$. 
This indicates that MM4 could be an unbound structure, perhaps a knot in the red-shifted outflow lobe driven by MM1, likely to disperse back into the ISM if internal pressures continue to dominate over gravity. On the other hand, MM4 could be on the cusp of boundedness at the lower end of our temperature range, possibly indicative of the early stages of a prestellar core.

\subsection{Chemistry of MM1 in context}
\label{sec:abundances}

\begin{figure*}
\centering
\includegraphics[scale=.568]{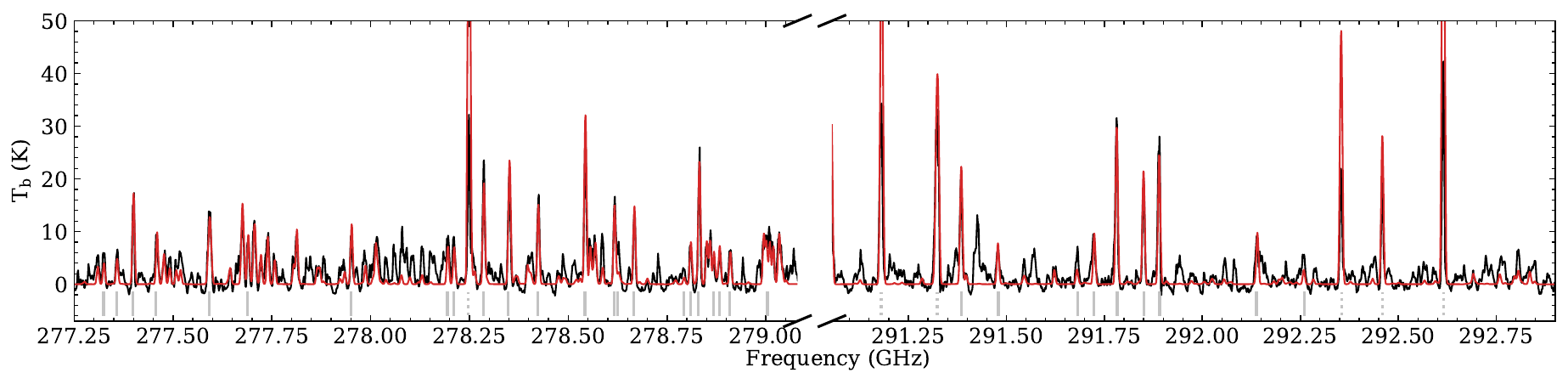}
\caption{Observed spectra towards the ALMA 1.05\,mm continuum peak (black) of MM1, overplotted with the combined highest-likelihood LTE synthetic spectra (red) for all COM species listed in Table~\ref{tab:chemistry}. Vertical grey lines mark the transitions identified in \S\ref{sec:lines}, with dotted lines indicating transitions mentioned in \S\ref{sec:abundances_estimation} that have overestimated brightness temperatures in the synthetic spectrum. }
\label{fig:other_COMs}
\end{figure*}

As discussed in Sections~\ref{sec:intro} and \ref{sec:lines}, the detection of a line forest towards MM1 with ALMA (see Figure~\ref{fig:spec6}) reveals a chemical richness that was lacking in the SMA observations of \cite{cyganowski11a}. We here estimate the abundances of these newly detected molecular species towards MM1, to see where MM1 sits in relation to other sources in the literature.

\subsubsection{Estimation of likely column densities and abundances}
\label{sec:abundances_estimation}

In Section\textbf{s}~\ref{sec:synthspec} and \ref{sec:synthspec_results}, we produced LTE synthetic spectra to evaluate the highest likelihood column density, temperature, centroid velocity and velocity width of CH$_{3}$OH towards MM1. However, the other detected species listed in Table~\ref{tab:lines} do not have enough identified transitions for the constraint of four free parameters, nor for the construction of a rotation diagram as in Section~\ref{sec:rotdiag}. We instead evaluate 
order of magnitude estimates for the column densities of these species (CH$_{3}$OCH$_{3}$, $^{13}$CH$_{3}$OH, CH$_{3}$CH$_{2}$OH, CH$_{3}$OCHO, OCS, H$_{2}$CO, CH$_{3}$CHO, H$_{2}$CS, CH$_{3}$CH$_{2}$CN, NH$_{2}$CHO, HC$_{3}$N and $^{13}$CS) at the ALMA 1.05\,mm continuum peak of MM1 
by keeping only the column density as a free parameter \citep[e.g.][]{csengeri19} in the {\sc emcee} sampler with {\sc weedspy\_mcmc}. The synthetic spectra for each molecular species are evaluated individually.  
As in Section~\ref{sec:synthspec_results}, the source size is fixed to the geometric mean of the ${\sim}278$\,GHz synthesised beam (${\sim}0.5\arcsec$). The temperature, centroid velocity and velocity width are set to the highest likelihood values evaluated at the ALMA 1.05\,mm continuum peak from the CH$_{3}$OH spectra (i.e. 162.7\,K, 59.7\,km\,s$^{-1}$ and 6.4\,km\,s$^{-1}$ respectively, see Figure~\ref{fig:corner}). 
As discussed in \paperone{}, for molecular lines associated with MM1 (i.e.\ not outflow-tracing) the extent of the line emission is generally consistent across species and with the extent of the 1.05\,mm continuum emission \citep[e.g.\ Fig.~3 of][]{williams22}.
For molecular species with transitions in multiple spectral windows, 
the synthetic spectra were evaluated in the spectral window with the most transitions. The resulting highest likelihood column density was then used to generate synthetic spectra for the other spectral windows. 

The highest likelihood column density of each species is listed in Table~\ref{tab:chemistry}, and Figure~\ref{fig:other_COMs} shows the combined LTE synthetic spectrum for the two wide-band spectral windows for all identified molecular species. On the whole the combined model reproduces the observed emission well. Some molecular lines appear in the synthetic spectrum that were not listed in Table~\ref{tab:lines} (or labelled in Figure~\ref{fig:spec6} or Figure~\ref{fig:other_COMs}), which is attributed to overlapping lines from the same and/or multiple species making their unequivocal identification in Section~\ref{sec:lines} difficult. 
Some of the identified transitions have their emission over-estimated. These include the CH$_{3}$OH($6_{1,5}-5_{1,4}$) and CH$_{3}$OH($10_{1,10}-9_{0,9}$) lines (with $\nu_{\mathrm{rest}}=292.67291$ and $292.51744$\,GHz respectively), the former being of an outflow-tracing nature (see Figure~\ref{fig:peakmaps}) with an apparently low integrated intensity in the rotation diagram of Figure~\ref{fig:rotdiag}(a), and the latter being the only transition identified in the tuning to be in the first torsionally excited state. 
The heavily blended H$_{2}$CO($4_{2,3}-3_{2,2}$) and CH$_{3}$OH($15_{1,0}-14_{2,0}$) lines (with $\nu_{\mathrm{rest}}=291.23777$ and $291.24057$\,GHz respectively) 
also have their intensity over-estimated, which we attribute to the line blending and the limitations of linearly combining independent models of different species.
The H$_{2}$CO($4_{2,2}-3_{2,1}$) line ($\nu_{\mathrm{rest}}=291.94807$\,GHz) also has its intensity over-estimated, attributed to this line also being heavily blended. Finally, the CH$_{3}$OCH$_{3}$($16_{1,16}-15_{0,15}$) line ($\nu_{\mathrm{rest}}=292.41225$\,GHz) is over-estimated despite all other CH$_{3}$OCH$_{3}$ lines in the tuning having their intensities well reproduced, and it being shown in \paperone{} to exhibit compact emission consistent with the other identified lines.

Fractional abundances of each species with respect to CH$_{3}$OH and H$_{2}$ are evaluated from the ratio of their respective column densities. We take the CH$_{3}$OH column density at the ALMA 1.05\,mm continuum peak from the synthetic spectra in Section~\ref{sec:synthspec_results}, and the H$_{2}$ column density found in Section~\ref{sec:low-mass} from the ALMA 1.05\,mm continuum. 
As listed in Table~\ref{tab:chemistry}, we report fractional abundances of the identified species ranging from $10^{-1}-10^{-4}$ with respect to CH$_{3}$OH, and $10^{-6}-10^{-9}$ with respect to H$_{2}$.

\begin{table}
\centering
\caption{Highest likelihood column densities of molecular species identified towards the ALMA 1.05\,mm continuum peak of MM1, and their abundances with respect to CH$_{3}$OH and H$_{2}$, arranged by decreasing column density.}
\label{tab:chemistry}
\setlength\tabcolsep{8.0pt}
\begin{tabular}{lccc}
\hline
Molecule                & N(X)                     & \multicolumn{2}{c}{Fractional abundance}                \\
(X)                     & (cm$^{-2}$)           & N(X) / N(CH$_{3}$OH)         & N(X) / N(H$_{2}$)                      \\ \hline
CH$_{3}$OH              & $2.2\times10^{18}$      & $10^{0}$                     & $3.3\times10^{-6}$       \\
CH$_{3}$OCH$_{3}$       & $4.3\times10^{17}$      & $2.0\times10^{-1}$           & $6.5\times10^{-7}$       \\   
$^{13}$CH$_{3}$OH       & $3.9\times10^{17}$      & $1.8\times10^{-1}$           & $5.9\times10^{-7}$       \\
CH$_{3}$OCHO            & $1.3\times10^{17}$      & $5.9\times10^{-2}$           & $2.0\times10^{-7}$       \\
CH$_{3}$CH$_{2}$OH      & $6.0\times10^{16}$      & $2.7\times10^{-2}$           & $9.1\times10^{-8}$       \\
OCS                     & $4.4\times10^{16}$      & $2.0\times10^{-2}$           & $6.7\times10^{-8}$       \\
H$_{2}$CO               & $3.0\times10^{16}$      & $1.4\times10^{-2}$           & $4.6\times10^{-8}$      \\
CH$_{3}$CHO             & $1.7\times10^{16}$      & $7.7\times10^{-3}$           & $2.6\times10^{-8}$      \\
H$_{2}$CS               & $1.6\times10^{16}$      & $7.2\times10^{-3}$           & $2.4\times10^{-8}$      \\
CH$_{3}$CH$_{2}$CN      & $1.0\times10^{16}$      & $4.6\times10^{-3}$           & $1.5\times10^{-8}$      \\
NH$_{2}$CHO             & $5.0\times10^{15}$      & $2.3\times10^{-3}$           & $7.6\times10^{-9}$       \\
HC$_{3}$N               & $4.4\times10^{15}$      & $2.0\times10^{-3}$           & $6.6\times10^{-9}$      \\ 
$^{13}$CS               & $1.5\times10^{15}$      & $6.8\times10^{-4}$           & $2.3\times10^{-9}$      \\
\hline
\end{tabular}
\begin{justify}
Column 1: Molecule name. Col. 2: Column density of each identified molecule at the ALMA 1.05\,mm continuum peak of MM1 (Table~\ref{tab:leaves}). Col. 3: Abundance of each molecule with respect to the $^{12}$CH$_{3}$OH column density at the ALMA 1.05\,mm continuum peak of MM1 ($2.2\times10^{18}$\,cm$^{-2}$; Section~\ref{sec:synthspec_results}). Col. 4: As for column 3, but for H$_{2}$ ($6.6\times10^{23}$\,cm$^{-2}$; Table~\ref{tab:cont_masses}).
\end{justify}
\end{table}

\subsubsection{Comparison to other YSOs observed at high resolution}

\begin{figure}
\centering
\includegraphics[scale=.77]{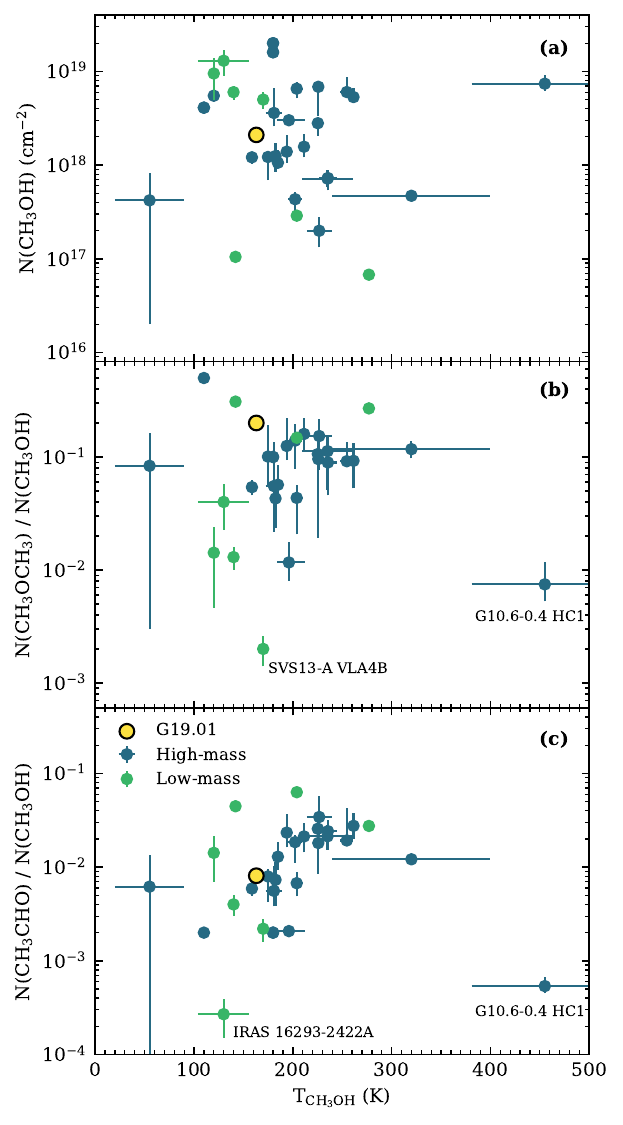}
\caption{(a) CH$_{3}$OH column density, (b) abundance of CH$_{3}$OCH$_{3}$ with respect to CH$_{3}$OH, and (c) abundance of CH$_{3}$CHO with respect to CH$_{3}$OH, all plotted against CH$_{3}$OH rotational temperature, of both low-mass (green) and high-mass (blue) sources from the literature \citep[inspired by Figs.7\&8 of][]{oberg14_faraday}. Our G19.01 data is plotted in yellow. All sources were observed with ALMA: the synthesised beams of the collated low- and high-mass literature observations range from $13-140$\,{\sc au} and $400-2600$\,{\sc au} respectively. High-mass sources were collated from \protect\cite{csengeri19}, \protect\cite{molet19}, \protect\cite{law21} and \protect\cite{baek22}, and low-mass sources were collated from \protect\cite{lee19}, \protect\cite{manigand20}, \protect\cite{bianchi22}, \protect\cite{chahine22} and \protect\cite{hsu22}. The G19.01 errors bars are smaller than the marker size, and not all literature sources have error bars.}
\label{fig:abundance_plot}
\end{figure}

To put the methanol column density and these fractional abundances into context, we compare them to values reported in the literature. 
Collated in Figure~\ref{fig:abundance_plot} are ALMA observations for both low-mass YSOs and high-mass hot cores that have reported observations of CH$_{3}$OH, CH$_{3}$OCH$_{3}$ and CH$_{3}$CHO, and that have rotational temperatures and CH$_{3}$OH column densities derived from the main CH$_{3}$OH isotopologue. Our MM1 data are overplotted for comparison. Compared to the other high-mass YSOs shown in Figure~\ref{fig:abundance_plot}, the CH$_{3}$OH column density, temperature and CH$_{3}$CHO fractional abundance of MM1 appear consistent, whilst MM1 may be one of the most abundant sources in CH$_{3}$OCH$_{3}$ (Figure~\ref{fig:abundance_plot}b).

It is also interesting to consider the low- and high-mass sources in Figure~\ref{fig:abundance_plot} in relation to each other. 
Of the high-mass sources shown, G10.6$-$0.4 hot core 1 \citep{law21} has the lowest CH$_{3}$OCH$_{3}$ and CH$_{3}$CHO fractional abundances (Figure~\ref{fig:abundance_plot}b, c), a result of its comparatively low CH$_{3}$OCH$_{3}$ and CH$_{3}$CHO column densities. Of the low-mass sources, IRAS\,16293-2422\,A \citep{manigand20} has the lowest CH$_{3}$CHO fractional abundance (Figure~\ref{fig:abundance_plot}c), and is noted to have more extended CH$_{3}$CHO emission morphology compared to the other COMs, whilst SVS13$-$A\,VLA4B \citep{bianchi22} exhibits the lowest CH$_{3}$OCH$_{3}$ abundance (Figure~\ref{fig:abundance_plot}b). 
However despite these outliers, the general parameter spaces occupied by the low- and high-mass sources in Figure~\ref{fig:abundance_plot} appear remarkably similar to each other.
Regarding the CH$_{3}$OH column density and temperature in particular, single-dish observations in the literature have previously reported a distinct separation in these parameters across mass scales. \cite{oberg14_faraday} for instance with their literature sample show low CH$_{3}$OH column density and temperature towards low-mass COM-emitting sources, and high CH$_{3}$OH column density and temperature towards high-mass COM-emitting sources. With beam dilution a potential contributing factor to this, \cite{oberg14_faraday} also collate a handful of resolved SMA observations, noting the distribution to be more continuous rather than separated across mass when including resolved observations. 
With the more recent ALMA observations collated here, this behaviour is even more pronounced, 
with the chemistry  of low- and high-mass sources appearing similar in this set of interferometric observations.
A similar trend has recently been noted by \cite{chen23}, who compare the abundance ratios of Oxygen-bearing COMs towards their ALMA sample of 14 high-mass sources with interferometric observations of 5 low-mass sources from the literature.
The \cite{chen23} sample includes G19.01--0.03 MM1, and their values for the CH$_{3}$OCH$_{3}$, CH$_{3}$OCHO and CH$_{3}$CHO  column densities towards MM1 -- derived using LTE synthetic spectra -- are within a factor of $2-4$ of the order of magnitude estimates in Table~\ref{tab:chemistry}.  We note that, for consistency, we do not include datapoints from \cite{chen23} in Figure~\ref{fig:abundance_plot} because they derive CH$_{3}$OH column densities from modelling of the CH$_{3}^{18}$OH isotopologue and an assumed $^{16}$O$/^{18}$O isotope ratio, and do not derive T$_{\rm CH_{3}OH}$ values for most (13/14) of the sources in their sample.

It is worth emphasising that comparisons of column densities are subject to a number of varying effects including beam sizes, source distance and beam dilution. The sources collated here, despite all being observed with ALMA, appear to trace different spatial scales across mass. The low-mass data probe physical scales of $13-140$\,{\sc au} (physical scale of the synthesised beam; median and standard deviation of 60 and 40\,{\sc au} respectively), a factor of up to two orders of magnitude smaller than the $400-2600$\,{\sc au} spatial scales (physical scale of the synthesised beam; median and standard deviation of 990 and 870\,{\sc au} respectively) probed by the high-mass data.
It is suggested by \cite{vangelder22} that methanol production could be less efficient towards MYSOs due to their warmer prestellar phase and/or shorter prestellar lifetimes. 
\cite{nazari22c} model the methanol emission towards different configuration high-mass sources. Comparing their results to a similar study of theirs towards low-mass sources \citep{nazari22b}, they find that optically thick millimetre dust, as well as disc shadowing causing a reduction in environment temperature, are more effective in lowering the methanol emission towards low-mass YSOs than they are towards MYSOs with high luminosity ($10^{4}-10^{5}$\,L$_{\odot}$). It is therefore further suggested by \cite{nazari22c} that other factors such as the presence of H{\sc ii} regions, could contribute to the lowering of methanol emission towards some MYSOs.  It is possible a combination of these factors could be contributing to the methanol column densities, temperature and fractional abundances appearing so similar across mass scales in Figure~\ref{fig:abundance_plot}. 
It may also be explained by considering the conditions under which these COMs are formed. Recent works have suggested that the chemical similarity between various sources may be explained by the COMs being formed under similar conditions \citep[e.g.][]{quenard18}. Across mass scales this would correspond to the ices of the pre-stellar phase \citep[e.g.][]{coletta20,nazari22b,chen23}.

In addition to CH$_{3}$OCH$_{3}$ and CH$_{3}$CHO, \cite{csengeri19} also report the fractional abundances of CH$_{3}$OCHO, CH$_{3}$CH$_{2}$OH and CH$_{3}$CH$_{2}$CN with respect to CH$_{3}$OH towards three positions around the high-mass hot core G328.2551--0.5321. In every instance, the corresponding molecular abundance reported here towards MM1 is of the same order as those towards G328.2551--0.5321. 
\cite{cyganowski11a} find a CH$_{3}$CN column density towards MM1 of $1.7\times10^{16}$\,cm$^{-2}$ with their SMA 1.3\,mm observations (with $32\times$ larger beam area than our ALMA data). 
They also find that the CH$_{3}$CN spectrum is best-fit by a model with a CH$_{3}$CN-emitting region of size 0.6\arcsec (equivalent to 2400\,AU at D\,$=$\,4\,kpc), intermediate between our ALMA beam (Table~\ref{tab:obs}) and the size of the MM1 1.05\,mm dust continuum from Table~\ref{tab:leaves}. 
\cite{nazari22b} present CH$_{3}$CN column densities towards 37 high-luminosity, potentially high-mass, protostars; 13 have column densities of a similar order to MM1 \citep{cyganowski11a}, whilst \cite{law21} reports a CH$_{3}$CN column density for G10.6 HC 1 an order of magnitude larger than MM1.

The comparatively weak, sparse COM emission towards MM1 in the SMA observations of \cite{cyganowski11a} led to the suggestion that MM1 was a relatively young source. Now, with a rich line forest detected with ALMA, and fractional abundances of COMs similar to other sources in the literature, it is clear that MM1 is not especially chemically young and in fact appears to be a typical hot core.

\subsubsection{$^{12}\mathrm{C}/^{13}\mathrm{C}$ isotope ratio}

With the identification of both the $^{12}$C and $^{13}$C isotopologues of CH$_{3}$OH, we can estimate the $^{12}\mathrm{C}/^{13}\mathrm{C}$ isotope ratio from the ratio of their column densities (see Table~\ref{tab:chemistry}). This is valid if the molecular transitions are optically thin, described by the same rotational temperature, and exhibit similar spatial emission extents \citep[e.g.][]{wirstrom11}.  
Towards the ALMA 1.05\,mm continuum peak of MM1, we estimate that $^{12}\mathrm{C}/^{13}\mathrm{C}\simeq6$. This is a factor of $\sim$7 lower than expected from the relations of \cite{wilson94} and \cite{milam05} for galactic molecular clouds (where $^{12}\mathrm{C}/^{13}\mathrm{C}\simeq39-46$ at the MM1 galactocentric distance of 4.4\,kpc). 

Though low-J transitions of CH$_{3}$OH are commonly optically thick \citep[e.g.][]{ginsburg17}, 5 out of the 6 lines included in our LTE line modelling (see \S\ref{sec:synthspec_results}) are relatively high-J transitions (see Table~\ref{tab:lines}). 
The estimated opacities of the 6 lines are $\tau~{\sim}~0.1$ to 0.6 based on the {\sc class} modelling (with an average of 0.3; see \S\ref{sec:synthspec_results}), and $\tau~{\sim}~0.4$ to 2.1 from the rotation diagram analysis (with an average of 1.0; see \S\ref{sec:rotdiag}).  
The opacity of the $^{13}$CH$_{3}$OH($3_{2,2}-4_{1,3}$) line reported by {\sc class} is $\tau=1.0$.
It is worth noting that the LTE modelling reproduces the observed emission of the highest-J and lowest-$\tau$ $^{12}$CH$_3$OH line very well (i.e. CH$_{3}$OH($23_{4,19}-22_{5,18}$) with $\tau=0.1$ and $0.4$ from {\sc class} and the rotation diagram respectively), and that the spatial extent of emission from this line is consistent with that of  $^{13}$CH$_{3}$OH($3_{2,2}-4_{1,3}$) (see Fig.3 of \paperone). 
To test for dependence on the assumed source size, we reran our \textsc{weedspy\_mcmc} analysis at the 1.05\,mm continuum peak for the case of unresolved emission (source size 0\farcs2, \S\ref{sec:rotdiag}) and re-estimated the $^{13}$CH$_3$OH column density as described in Section~\ref{sec:abundances_estimation} using the resulting parameters.   We emphasise that the best-fitting model with this smaller source size is a poorer representation of the observed $^{12}$CH$_3$OH emission than the model described in Sections~\ref{sec:synthspec_results}-\ref{sec:synthspec_images}, and we use it only to check the effect of the assumed source size on the $^{12}\mathrm{C}/^{13}\mathrm{C}$ ratio. From the models with a 0\farcs2 source size, we derive a $^{12}\mathrm{C}/^{13}\mathrm{C}$ ratio of $\sim$7.5, indicating that the result of low $^{12}\mathrm{C}/^{13}\mathrm{C}$ is robust to the assumption of resolved or unresolved emission. 

There are suggestions in the literature that $^{12}\mathrm{C}/^{13}\mathrm{C}$ may be generally lower on the smaller scales of high-mass hot cores \citep[e.g.\ IRAS~20126+4104, G31.41$-$0.31, AFGL~4176, W3~IRS4 and G10.6$-$0.4;][respectively]{palau17, beltran18, bogelund19, mottram20, law21} and low-mass hot corinos \citep{hsu22} than on the larger scale of molecular clouds. 
Proposed chemical explanations include cold temperature enhancement of $^{13}$C in H$_{2}$CO formation on dust grains \citep{mottram20} and the destruction of HC$_{3}$N in chemical reactions \citep{bogelund19}.  
The possible contribution of optical depth effects has also been considered \citep{beltran18,bogelund19,law21}.
\cite{beltran18} and \cite{bogelund19} conduct synthetic analyses of CH$_{3}$CN and HC$_{3}$N emission respectively (taking similar approaches to that detailed here towards G19.01), and comment that optically thick emission could 
artificially lower their observed $^{12}\mathrm{C}/^{13}\mathrm{C}$ ratios.  \cite{law21} conclude that optical depth effects are unlikely to drive their results as $\tau~{\sim}~0.1-0.4$ for their CH$_3$OH lines (derived using a rotation diagram analysis), but note they cannot definitively discount optical depth effects in unresolved structures contributing to their results.

Similarly to \citet{law21}, our derived line opacities are not suggestive of optical depth being the primary explanation for the low $^{12}\mathrm{C}/^{13}\mathrm{C}$ ratio observed in G19.01, but we cannot rule out optical depth effects contributing to this result. 
We emphasise that our observations provide only a tentative addition to the emerging picture of low $^{12}\mathrm{C}/^{13}\mathrm{C}$ isotope ratios towards hot cores.
Observations of additional and optically thin transitions of the $^{12}$CH$_{3}$OH and $^{13}$CH$_{3}$OH isotopologues, as well as the isotopologues of other molecular species, would be required for confirmation.

\section{Conclusions}
\label{sec:conclusions}

In this paper (Paper II), we have presented a study of the physical properties and chemistry of the high-mass (proto)star G19.01--0.03 MM1 and its environment, using sub-arcsec-resolution ALMA 1.05\,mm and VLA 1.21\,cm data. Our main findings are as follows:

\medskip

(i) A (sub)millimetre forest of molecular line emission is observed towards MM1. We analyse 47 lines from 11 different species (43 of which were identified in our ALMA 1.875\,GHz wide-bands alone), including COMs such as CH$_{3}$OCHO, CH$_{3}$CH$_{2}$CN, CH$_{3}$CH$_{2}$OH, CH$_{3}$OCH$_{3}$, NH$_{2}$CHO and CH$_{3}$OH. 

\smallskip

(ii) A Bayesian analysis (using our publicly available \textsc{weedspy\_mcmc} scripts) of CH$_{3}$OH LTE synthetic spectra towards the MM1 dust peak returns highest likelihood column density and temperature of $(2.22\pm0.01)\times10^{18}$\,cm$^{-2}$ and $162.7\substack{+0.3 \\ -0.5}$\,K respectively. These are consistent with high-mass hot core sources from the literature, and with values found from our opacity-corrected rotation diagram analysis of $(2.0\pm0.4)\times10^{18}$\,cm$^{-2}$ and $166\pm9$\,K respectively. 

\smallskip

(iii) The peak CH$_{3}$OH temperature ($165.5\pm0.6$\,K) is offset from the MM1 dust peak by 0.22$\arcsec\sim880$\,{\sc au}. The morphology of the region of elevated temperature is aligned with the VLA 5.01\,cm continuum emission presented in \paperone{}.

\smallskip

(iv) We report abundances of all identified molecular species (CH$_{3}$OCH$_{3}$, $^{13}$CH$_{3}$OH, CH$_{3}$CH$_{2}$OH, CH$_{3}$OCHO, OCS, H$_{2}$CO, CH$_{3}$CHO, H$_{2}$CS, CH$_{3}$CH$_{2}$CN, NH$_{2}$CHO, HC$_{3}$N and $^{13}$CS) of $10^{-1}-10^{-4}$ and $10^{-6}-10^{-9}$ with respect to CH$_{3}$OH and H$_{2}$ respectively, consistent with high-mass hot core sources from the literature.

\smallskip

(v) The known bipolar molecular outflow driven by MM1 is traced by thermal CH$_{3}$OH and H$_{2}$CO emission and by newly-identified  NH$_3$(3,3) and 278.3 GHz Class~I CH$_3$OH maser candidates, strengthening the connection of these types of masers with outflows driven by MYSOs. 
We identify a total of 50 $>$5$\sigma$ NH$_{3}$(3,3) emission spots across 8 compact emission groups, which are in the outer lobes of the outflow and spatially and kinematically coincident with 44\,GHz Class~I CH$_{3}$OH masers.
Candidate 25\,GHz CH$_{3}$OH 5(2,3)-5(1,4) maser emission is also detected towards the outflow, offset from MM1 by 6$\arcsec$\,$\sim$\,24,000\,{\sc au}.  

\smallskip

(vi) Four new millimetre continuum sources (MM2...MM5) are detected between $0.03 - 0.12$\,pc from MM1, but all are undetected in NH$_{3}$($J$=$K$=1,2,3,5,6,7) and 25\,GHz CH$_{3}$OH emission.  
Two of these sources, MM2 and MM4, are associated with $>5\sigma$ DCN(4--3) emission, a tracer of cold and dense gas.

\smallskip

(vii) The median mass of MM2..MM5 (assuming isothermal dust emission) is 1.0$-$2.0\,M$_{\odot}$ for a temperature range of $24-15$\,K, and their median radius is $\sim$\,2000\,{\sc au}.
Since none of these sources are associated with any masers or have any sign of outflows, they are candidate low-mass pre-stellar companions to MM1.

\medskip

In all, our results place G19.01--0.03 MM1 as a typical high-mass hot core source, with four low-mass, potentially pre-stellar, cores within a projected linear distance of 0.12\,pc. 
Both these findings are in contrast with the results of previous, lower-resolution SMA observations \citep{cyganowski11a}, which identified G19.01--0.03 MM1 as a candidate for isolated high-mass star formation, potentially at a very early stage due to its lack of chemical complexity. 
Our new results show that G19.01--0.03 MM1 is in fact not especially chemically young, and that its physical and chemical properties are typical of other high-mass hot core sources in the literature.
Intriguingly, our analysis of the ALMA CH$_{3}$OH lines reveals a region of elevated CH$_{3}$OH temperature that is aligned with VLA 5.01\,cm continuum emission attributed to free-free emission, which is offset from the thermal dust emission peak (as discussed in detail in \paperone{}).  This positional coincidence is tentative support for the possibility, suggested in \paperone{}, that the 5.01\,cm continuum might be associated with an unresolved high-mass binary companion to MM1. 
Our ALMA Cycle 6 follow-up study, tuned to probe $\sim$0.09\arcsec$\sim$360\,{\sc au} scales, will be a next step in testing this possibility with higher angular resolution observations.

\section*{Acknowledgements}

We thank the anonymous referee for their constructive report and comments that helped improve the quality of this paper. GMW and CJC thank Ian Bonnell for helpful discussions. GMW acknowledges support from the UK's Science and Technology Facilities Council (STFC) under ST$/$W00125X$/$1.  
CJC acknowledges support from the UK's STFC under ST$/$M001296$/$1. 
PN acknowledges support by grant 618.000.001 from the Dutch Research Council (NWO) and support by the Danish National Research Foundation through the Center of Excellence "InterCat" (Grant agreement no.: DNRF150).
This research has made use of: NASA's Astrophysics Data System Bibliographic Services, {\sc gildas} (\href{https://www.iram.fr/IRAMFR/GILDAS}{https://www.iram.fr/IRAMFR/GILDAS}) and {\sc python} packages {\sc astropy} \citep{astropy}, {\sc astrodendro} \citep{rosolowsky08dendro}, {\sc aplpy} (\href{http://aplpy.github.com}{http://aplpy.github.com}), {\sc cmocean} \citep{cmocean}, {\sc corner} \citep{corner}, {\sc emcee} \citep{emcee},  {\sc matplotlib} \citep{matplotlib}, {\sc numpy} \citep{numpy}, {\sc pandas} \citep{pandas2010}, {\sc scipy} \citep{scipy}, {\sc scitkit-image} \citep{scikit-image}, and {\sc analysisUtils} \citep{au2023}.
This paper makes use of the following ALMA data: ADS$/$JAO.ALMA$\#$2013.1.00812.S. ALMA is a partnership of ESO (representing its member states), NSF (USA) and NINS (Japan), together with NRC (Canada), NSC and ASIAA (Taiwan), and KASI (Republic of Korea), in cooperation with the Republic of Chile. The Joint ALMA Observatory is operated by ESO, AUI/NRAO and NAOJ. The National Radio Astronomy Observatory is a facility of the National Science Foundation operated under cooperative agreement by Associated Universities, Inc.
For the purposes of open access, the author has applied a Creative Commons Attribution (CC BY) licence to any Author Accepted Manuscript version arising.

\section*{Data Availability}

The ALMA 1.05\,mm dust continuum, VLA 1.21 and 5.01\,cm continuum images, and the line cubes for the two broad (1.875\,GHz bandwidth) ALMA spectral windows are available at \href{http://dx.doi.org/10.5281/zenodo.8059531}{doi:10.5281/zenodo.8059531}. These data underlie both this article and \cite{williams22}. The scripts used to conduct the Bayesian analysis of LTE synthetic spectra are packaged together in a reusable form we call \textsc{weedspy\_mcmc}, and are made freely available at \url{https://github.com/gwen-williams/WeedsPy\_MCMC}.



\bibliographystyle{mnras}
\bibliography{bib} 




\appendix

\section{N$_{2}$H$^{+}$(3--2) absorption towards MM1}
\label{appendix:n2hp}

\begin{figure}
\centering
\includegraphics[scale=.75]{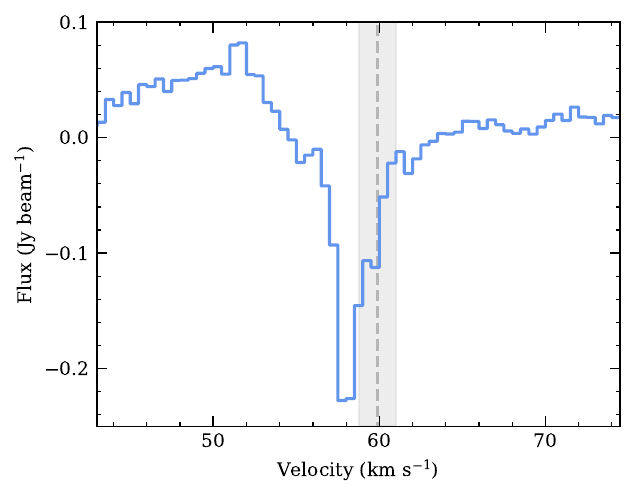}
\caption{N$_{2}$H$^{+}$(3--2) spectrum towards the ALMA 1.05\,mm continuum peak of MM1. The systemic velocity of MM1 \citep[$59.9\pm1.1$\,km\,s$^{-1}$;][]{cyganowski11a} is marked by the vertical grey dashed line (with the uncertainty represented by the shaded region).}
\label{fig:n2hp_spec}
\end{figure}

Figure~\ref{fig:n2hp_spec} shows the N$_{2}$H$^{+}$(3--2) spectrum towards the MM1 ALMA 1.05\,mm continuum peak, extracted from the continuum-subtracted line image cube.  As shown in Figure~\ref{fig:n2hp_spec}, the deepest absorption (at $\sim$58\,km\,s$^{-1}$) is blue-shifted relative to the systemic velocity of MM1 measured from its hot-core line emission \citep[$59.9\pm1.1$\,km\,s$^{-1}$;][]{cyganowski11a}.

\section{Convergence of walkers in \texttt{weedspy$\_$mcmc} analysis of LTE synthetic spectra}
\label{appendix:emcee}

In Section~\ref{sec:synthspec}, we used the \textsc{python} package \textsc{emcee} to sample the parameter space of the LTE synthetic spectra produced by the Weeds package of the \textsc{class} software. Here, we discuss how we heuristically assessed algorithm parameters.

\begin{figure}
\centering
\includegraphics[scale=.75]{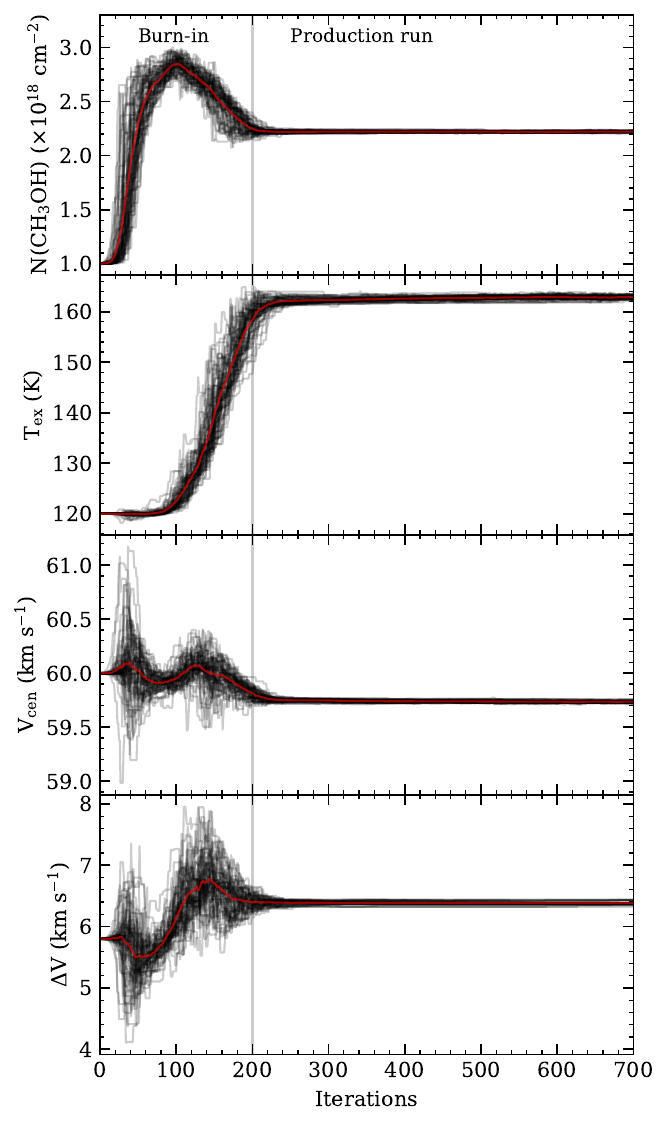}
\caption{Trace plots of the chains of 60 walkers across the four free parameters of the analysis (column density, temperature, centroid velocity and velocity width), for the total number of 700 iterations, evaluated for the spectrum towards the peak position of the ALMA 1.05\,mm continuum emission of MM1. The running mean is overplotted in red in each panel. The sampler was allowed to ``burn-in'' for the first 200 iterations (marked by the vertical grey line), after which a further 500 ``production'' iterations were run.}
\label{fig:trace}
\end{figure}

We ran a number of tests on the spectrum at the peak ALMA 1.05\,mm continuum pixel of MM1, with varying numbers of walkers and iterations. Figure~\ref{fig:trace} shows an example trace plot for the chains (i.e. walked paths) of 60 initiated walkers, allowed to walk for a total of 700 iterations each. We find that 200 burn-in iterations is sufficient for the walkers to locate the regions of highest probability in the parameter space. Since our analysis towards MM1 requires running this analysis on $\sim120$ observed spectra, our choice of production run iterations is mostly motivated by computational time constraints. For 500 production run iterations, the analysis for one spectrum takes 30\,minutes to complete. To assess whether this is appropriate or not, we calculate the acceptance fraction ($a_{f}$) and the integrated autocorrrelation time ($\tau_{ac}$). The acceptance fraction describes the fraction of iterations that were accepted, whilst the autocorrelation time describes the number of iterations the sampler takes to draw independent samples (i.e. effectively the number of iterations it takes to forget where the walkers were initialised). Good samplers are generally quoted to have $a_{f}=0.2-0.5$ \citep[e.g.][]{emcee}, whilst the lower the autocorrelation time the better. We calculate a mean $\tau_{ac}$ over 500 production iterations of 45, 50, 60 and 35 iterations for the four free parameters of the model respectively. \cite{emcee} however recommend that the number of production iterations should be at least $50\tau_{ac}$. This would require 3000 production iterations, a six-fold increase on the 500 iterations shown in Figure~\ref{fig:trace}, resulting in a computational time cost across all 120 spectra of more than 2\,weeks. This was unrealistic for our analysis. However, it is encouraging that the burn-in iterations are seen to be a number of multiples of the estimated $\tau_{ac}$ (between 3 and 6), and that the production iterations successfully focus on the region of highest probability (as seen in Figure~\ref{fig:trace}).  As convergence of such a sampler is always a non-trivial metric to estimate, and given that our estimated $a_{f}=0.21$ is just within the typically quoted acceptable range from the literature, we consider that for all intents and purposes these chains are ``converged'', with final highest likelihood results that are valid within the confidence intervals of the normally distributed posterior distributions shown in Figure~\ref{fig:corner}. Only the posterior distribution for the velocity width in Figure~\ref{fig:corner} does not appear normally distributed, as mentioned in Section~\ref{sec:synthspec}. Though the statistical error on the velocity width is significantly lower than the spectral resolution of the data, this motivated a final test where we initialised the 60 walkers with a different random number seed. This yielded very similar chain traces and posterior distributions, giving confidence that the final result is independent of initial walker position.

\bsp	
\label{lastpage}
\end{document}